\documentclass{emulateapj}
 
\shorttitle{Doppler Monitoring of five {\it K2} Transiting Planetary Systems}
\shortauthors{Dai et al.}

\usepackage{amsmath}
\usepackage{epstopdf}
\usepackage{apjfonts}
\usepackage{xspace}
\usepackage{hyperref} 
\usepackage{bm}
\usepackage{graphicx}

\begin{document}


\title{Doppler Monitoring of five {\it K2} transiting planetary systems}


\author{
Fei\ Dai\altaffilmark{1$\dagger$},
Joshua N.\ Winn\altaffilmark{1},
Simon\ Albrecht\altaffilmark{2},  
Pamela\ Arriagada\altaffilmark{3},
Allyson Bieryla\altaffilmark{3},
R.~Paul\ Butler\altaffilmark{4},
Jeffrey D.\ Crane\altaffilmark{5},
Teruyuki Hirano\altaffilmark{7}, 
John Asher\ Johnson\altaffilmark{4}, 
Amanda Kiilerich\altaffilmark{2},
David W.\ Latham\altaffilmark{4},
Norio\ Narita\altaffilmark{8,9,10}, 
Grzegorz Nowak\altaffilmark{11,12}, 
Enric Palle\altaffilmark{11,12},
Ignasi\ Ribas\altaffilmark{13},
Leslie A. Rogers\altaffilmark{14,15},
Roberto\ Sanchis-Ojeda\altaffilmark{16,15}, 
Stephen A.\ Shectman\altaffilmark{5}, 
Johanna K.\ Teske\altaffilmark{4,5}, 
Ian B.\ Thompson\altaffilmark{5}, 
Vincent\ Van Eylen\altaffilmark{2,1},
Andrew\ Vanderburg\altaffilmark{4},
Robert A.\ Wittenmyer\altaffilmark{17,18},
Liang\ Yu\altaffilmark{1}
}

\altaffiltext{1}{Department of Physics, and Kavli Institute for Astrophysics and Space Research,
  Massachusetts Institute of Technology, Cambridge, MA 02139, USA }
\altaffiltext{$\dagger$}{\url{fd284@mit.edu}}
\altaffiltext{2}{Stellar Astrophysics Centre, Department of Physics and Astronomy, Aarhus University, Ny Munkegade 120, DK-8000 Aarhus C, Denmark}
\altaffiltext{3}{Harvard-Smithsonian Center for Astrophysics, 60 Garden Street, Cambridge, MA 02138, USA}
\altaffiltext{4}{Carnegie Institution of Washington, Department of Terrestrial Magnetism, 5241 Broad Branch Road, NW, Washington DC, 20015-1305, USA}
\altaffiltext{5}{The Observatories of the Carnegie Institution of Washington, 813 Santa Barbara Street, Pasadena, CA 91101, USA}
\altaffiltext{6}{Institut d'Astrophysique et de Géophysique, Université de Liège, Allée du 6 Août 17, Sart Tilman, 4000 Liège 1, Belgium}
\altaffiltext{7}{Department of Earth and Planetary Sciences, Tokyo Institute of Technology, 2-12-1 Ookayama, Meguro-ku, Tokyo 152-8551, Japan}
\altaffiltext{8}{Astrobiology Center, National Institutes of Natural Sciences, 2-21-1 Osawa, Mitaka, Tokyo 181-8588, Japan}
\altaffiltext{9}{National Astronomical Observatory of Japan, 2-21-1 Osawa, Mitaka, Tokyo 181-8588, Japan}
\altaffiltext{10}{SOKENDAI (The Graduate University for Advanced Studies), 2-21-1 Osawa, Mitaka, Tokyo 181-8588, Japan}
\altaffiltext{11}{Instituto de Astrofísica de Canarias (IAC), 38205 La Laguna, Tenerife, Spain}
\altaffiltext{12}{Departamento de Astrofísica, Universidad de La Laguna (ULL), 38206 La Laguna, Tenerife, Spain}
\altaffiltext{13}{Institut de Ci\`encies de l'Espai (CSIC-IEEC), Carrer de Can Magrans, Campus UAB, 08193 Bellaterra, Spain}
\altaffiltext{14}{Department of Earth and Planetary Science, University of California, Berkeley, CA 94720, USA}
\altaffiltext{15}{NASA Sagan Fellow}
\altaffiltext{16}{Department of Astronomy, University of California, Berkeley, CA 94720, USA}
\altaffiltext{17}{School of Physics and Australian Centre for Astrobiology, University of New South Wales, Sydney 2052, Australia}
\altaffiltext{18}{Computational Engineering and Science Research Centre, University of Southern Queensland, Toowoomba, Queensland 4350, Australia}

\begin{abstract}

  \noindent In an effort to measure the masses of planets discovered by the NASA {\it K2} mission, we have conducted precise Doppler observations of five stars with transiting planets. We present the results of a joint analysis of these new data and previously published Doppler data. The first star, an M dwarf known as K2-3 or EPIC~201367065, has three transiting planets (``b'', with radius $2.1~R_{\oplus}$; ``c'', $1.7~R_{\oplus}$; and ``d'', $1.5~R_{\oplus}$). Our analysis leads to the mass constraints: $M_{b}=8.1^{+2.0}_{-1.9}~M_{\oplus}$ and $M_{c}$ < $ 4.2~M_{\oplus}$~(95\%~conf.). The mass of planet d is poorly constrained because its orbital period is close to the stellar rotation period, making it difficult to disentangle the planetary signal from spurious Doppler shifts due to stellar activity. The second star, a G dwarf known as K2-19 or EPIC~201505350, has two planets (``b'', $7.7~R_{\oplus}$; and ``c'', $4.9~R_{\oplus}$) in a 3:2 mean-motion resonance, as well as a shorter-period planet (``d'', $1.1~R_{\oplus}$). We find $M_{b}$= $28.5^{+5.4}_{-5.0} ~M_{\oplus}$, $M_{c}$= $25.6^{+7.1}_{-7.1} ~M_{\oplus}$ and $M_{d}$ < $14.0~M_{\oplus} $~(95\%~conf.). The third star, a G dwarf known as K2-24 or EPIC~203771098, hosts two transiting planets (``b'', $5.7~R_{\oplus}$; and ``c'', $7.8~R_{\oplus}$) with orbital periods in a nearly 2:1 ratio. We find $M_{b}$= $19.8^{+4.5}_{-4.4} ~M_{\oplus}$ and $M_{c}$ = $26.0^{+5.8}_{-6.1}~M_{\oplus}$. The fourth star, a G dwarf known as EPIC~204129699, hosts a hot Jupiter for which we measured the mass to be $1.857 ^{+0.081}_{-0.081}~M_{\text{Jup}}$. The fifth star, a G dwarf known as EPIC~205071984, contains three transiting planets (``b'', $5.4~R_{\oplus}$; ``c'', $3.5~R_{\oplus}$; and ``d'', $3.8~R_{\oplus}$), the outer two of which have a nearly 2:1 period ratio.
We find $M_{b}$= $21.1^{+5.9}_{-5.9} ~M_{\oplus}$, $M_{c}$ < $8.1~M_{\oplus}$~(95\%~conf.) and $M_{d}$ < $35~M_{\oplus}$~(95\%~conf.).

\end{abstract}

\keywords{planetary systems - planets and satellites: composition -
  stars: individual (K2-3 (EPIC~201367065), K2-19 (EPIC~201505350), EPIC~204129699, K2-24 (EPIC~203771098), EPIC~205071984) - techniques: radial velocities}

\section{Introduction}

The characterization of a planet starts with the measurement of its mass and radius. Without these two quantities, we cannot even begin to answer the most basic questions about the planet's internal structure, atmospheric composition and formation history. Of particular interest are planets with radii in the range of 1--4~$R_{\oplus}$, known as ``super-Earths'' or ``sub-Neptunes.'' Despite being the most frequently occurring exoplanets within 1~AU of solar-type stars \citep[e.g.][]{Petigura2013,Howard2012,Fressin2013}, they have no known counterparts in our solar system. Moreover, traditional core accretion models have trouble explaining why these planets did not undergo runaway accretion that would have led to the formation of gas giants \citep{Mizuno1980,Rafikov,Bodenheimer,Inamdar,LeeChiang2015}. Detailed characterization of these planets will shed light on these mysteries.

Only a few dozen Doppler mass measurements of planets smaller than Neptune have been reported \citep[see, e.g.,][]{Marcy2014,Howard2013,Pepe2013,Dressing2015,Motalebi2015, Gettel2016,Weiss2016}. The number has been limited because of the relative faintness of the host stars of the known transiting planets. The ongoing NASA {\it K2} mission \citep{Howard2014} is gradually providing a larger sample of stars which are bright enough to be amenable to precise Doppler follow-up observations. In this paper, we present new Doppler observations using three different high-resolution spectrographs. By combining these data with previously reported Doppler data \citep{Almenara2015,Petigura2016,Grziwa2015}, we placed mass constraints on the planets in five {\it K2} systems: K2-3, K2-19, K2-24, EPIC$~204129699$ and EPIC$~205071984$.

This paper is organized as follows. In Section 2, we present our transit candidate search pipeline and follow-up target selection which led us to the five {\it K2} systems. Section 3 summarizes the instruments and the details of our Doppler observations. Section 4 describes the methods we used to analyze the Doppler observations. Section 5 contains the results of our analysis for each planet candidate host. Section 6 discusses the implications of our results in a broader context.

\section{K2 Photometry and Target Selection}

The five systems presented in this paper have all been reported previously. They were also independently identified by the {\it K2} collaboration in which many of us participate, which is known as ESPRINT (``Equipo de Seguimiento de Planetas Rocosos INterpretando sus Tr\'{a}nsitos'' or the ``Follow-up team of rocky planets via the interpretation of their transits''). The production of calibrated and detrended {\it K2} light curves by the ESPRINT collaboration was described by \citet{Sanchis2015}.  The light curves were searched for transiting planet candidates with two algorithms. We employed the Box-Least-Squares routine \citep{Kovacs2002,Jenkins2010} using the optimal frequency sampling described by \citet{Ofir2014} and \citet{Vanderburg2016}. We also employed a Fast-Fourier-Transform method \citep{Sanchis2014} suitable for detecting short-period planet candidates (< 1 day). We then removed false positives by looking for the alternating eclipse depths associated with the primary and secondary eclipses of eclipsing binaries, and secondary eclipses deep enough that they could only be caused by a secondary star rather than a planet.

Candidates that passed these initial tests were then selected for Doppler follow-up observations based on their scientific interest and  measurement feasibility. Regarding scientific interest, we preferentially selected systems containing sub-Neptunes ($R<4~R_{\oplus}$), because their internal compositions and formation pathways are poorly understood. In order to estimate the planetary radius, we combined the measured transit depth with the stellar parameters derived from broadband photometries. Systems with multiple transiting planet candidates, especially those close to mean-motion resonances, were given higher priority in our selection process.  Multi-candidate systems are less likely to be false positives \citep{Latham2011,Lissauer2011multi} and transit timing variation (TTV) analysis of these systems may unveil the orbital configurations and offer an independent way of measuring masses. With the consideration of measurement feasibility, we only selected targets brighter than $V\approx 13$. Another factor affecting the feasibility of Doppler mass measurements is the level of stellar variability of the host star. Stellar activity such as spots and plages introduces radial-velocity perturbations on a characteristic timescale of the stellar rotation period. We tried to anticipate the level of rotation-induced radial velocity perturbations using the approximation:
\begin{equation}
\delta RV_{\text{\rm rot}} = \frac{2 \pi R_\star} {P_{\text{rot}}} \times \frac{\delta F}{F}
\end{equation}
where $\delta F/F$ is the observed fractional photometric variation and
$P_{\rm rot}$ is the rotation period, both of which were estimated from the {\it K2} data. We did not pursue systems for which $\delta RV_{\rm rot}$ was significantly higher than the anticipated Doppler signal. Using these criteria, we selected a few targets per {\it K2} campaign
for Doppler follow-up observations.

\section{Doppler Observations }
The Doppler observations presented in this paper were obtained with three spectrographs: the High Accuracy Radial-velocity Planet Searcher (HARPS) at the ESO La Silla 3.6m telescope, the Carnegie Planet Finder Spectrograph (PFS) on the 6.5m Magellan/Clay Telescope at Las Campanas Observatory in Chile, and the Tillinghast Reflector Echelle Spectrograph (TRES) on the 1.5m Tillinghast telescope at the Smithsonian Astrophysical Observatory's Fred L.\ Whipple Observatory on Mt.\ Hopkins in Arizona.

HARPS is an \'{e}chelle spectrograph that employs the simultaneous-reference method to achieve precise Doppler observations with long-term stability \citep{Mayor2003}. The spectral coverage of HARPS spans 378-691~nm and it has a spectral resolution of $\approx$115,000. The instrument is sealed in a vacuum container to maintain stability against temperature and pressure fluctuations. We used the Fabry-Perot etalon for simultaneous wavelength calibration. The exposure times ranged from 20-45 min. The signal-to-noise ratio (SNR) of the stellar spectra obtained near a wavelength range of 5500~\AA~ranged from 20-50. Radial velocities and uncertainties for our targets were derived using the standard HARPS Data Reduction Software.

PFS employs an iodine gas cell to superimpose well-characterized absorption features onto the stellar spectrum. The iodine absorption lines help to establish the wavelength scale and instrumental profile \citep{Crane2010}. The detector was read out in the standard $2 \times 2$ binned mode. Exposure times ranged from 15-20 min, giving a SNR of 50-140~pixel$^{-1}$ and a resolution of about 76,000 in the vicinity of the iodine absorption lines. An additional iodine-free spectrum with higher resolution and higher SNR was obtained for each star, to serve as a template spectrum for the Doppler analysis. The relative radial velocities were extracted from the spectrum using the techniques of \citet{Butler1996}. The internal measurement uncertainties (ranging from 2.5-4~m~s$^{-1}$) were estimated based on the scatter in the results to fitting individual 2~\AA~sections of the spectrum. 
 
TRES is a fiber-fed \'{e}chelle spectrograph with resolving power of 44,000 and wavelength coverage 390-910~nm.  Wavelength calibration is achieved using exposures of a Thorium-Argon hollow cathode lamp through the science fiber before and after each observation. The relative velocities reported for EPIC~204129699 in this paper were derived by cross-correlating the individual observations against the strongest observation, which thus defines the velocity zero point. The mean SNR per resolution element for the ten observations reported here was 40, the average exposure time was 10 minutes, and the average internal error estimate, derived from the scatter of the velocities for the individual \'{e}chelle orders in each observation, was 15~m~s$^{-1}$.  Some of the orders were not included in the velocity determinations, either because the orders were contaminated by telluric lines introduced by the Earth's atmosphere, or the exposure level in the order was too weak, typically at the shortest wavelengths.

\begin{figure}
\begin{center}
\includegraphics[width= 1.1\columnwidth]{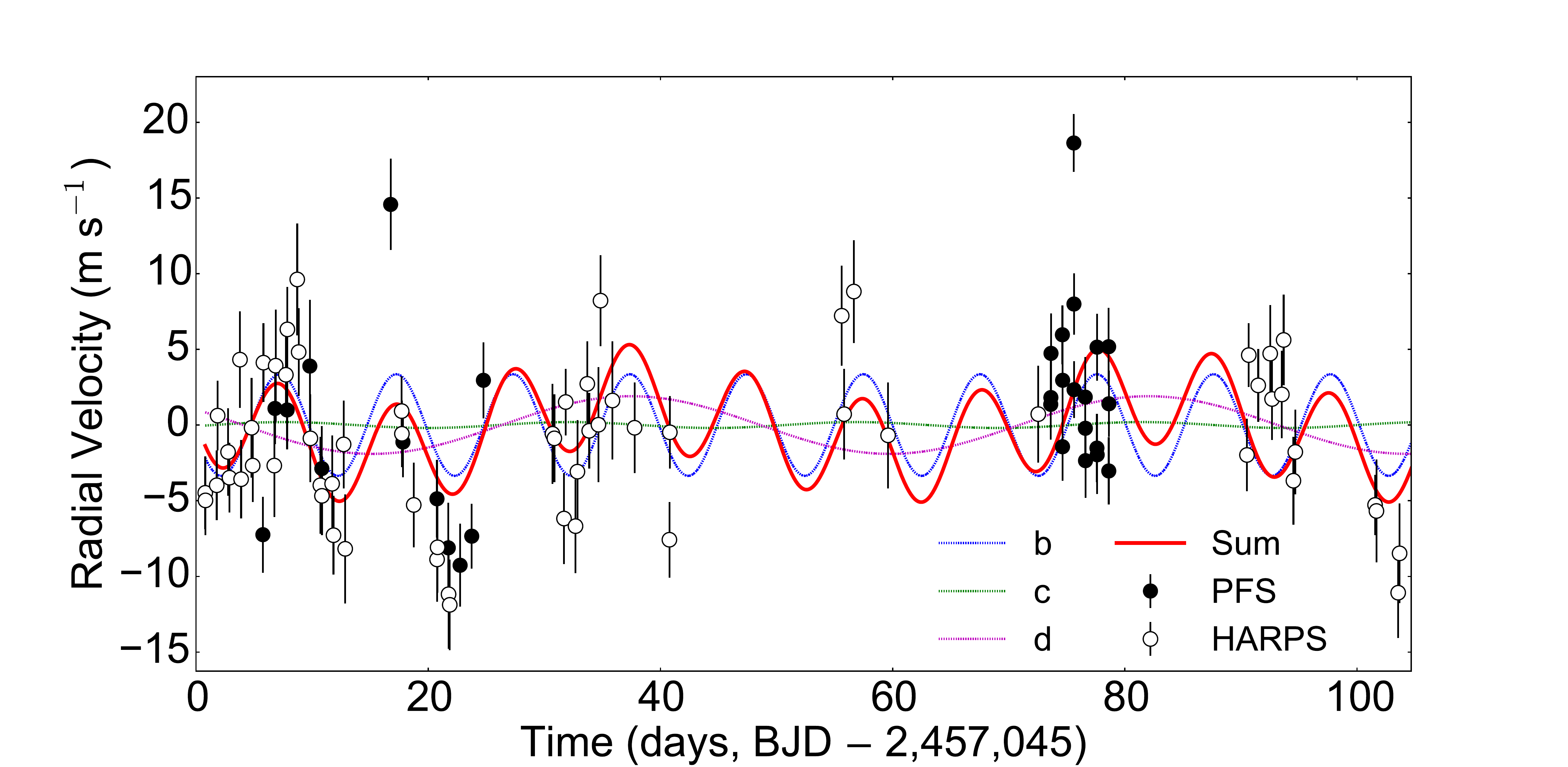}
\caption{Measured radial velocity of K2-3 (open circles are HARPS data from \citet{Almenara2015}; black circles are new PFS data) and the best-fitting model (red line) assuming circular orbits. The other colored lines show the contributions to the model curve from individual planets.}
\label{201367065rvplot}
\end{center}
\end{figure}

\begin{figure}
\begin{center}
\includegraphics[width= 0.9\columnwidth]{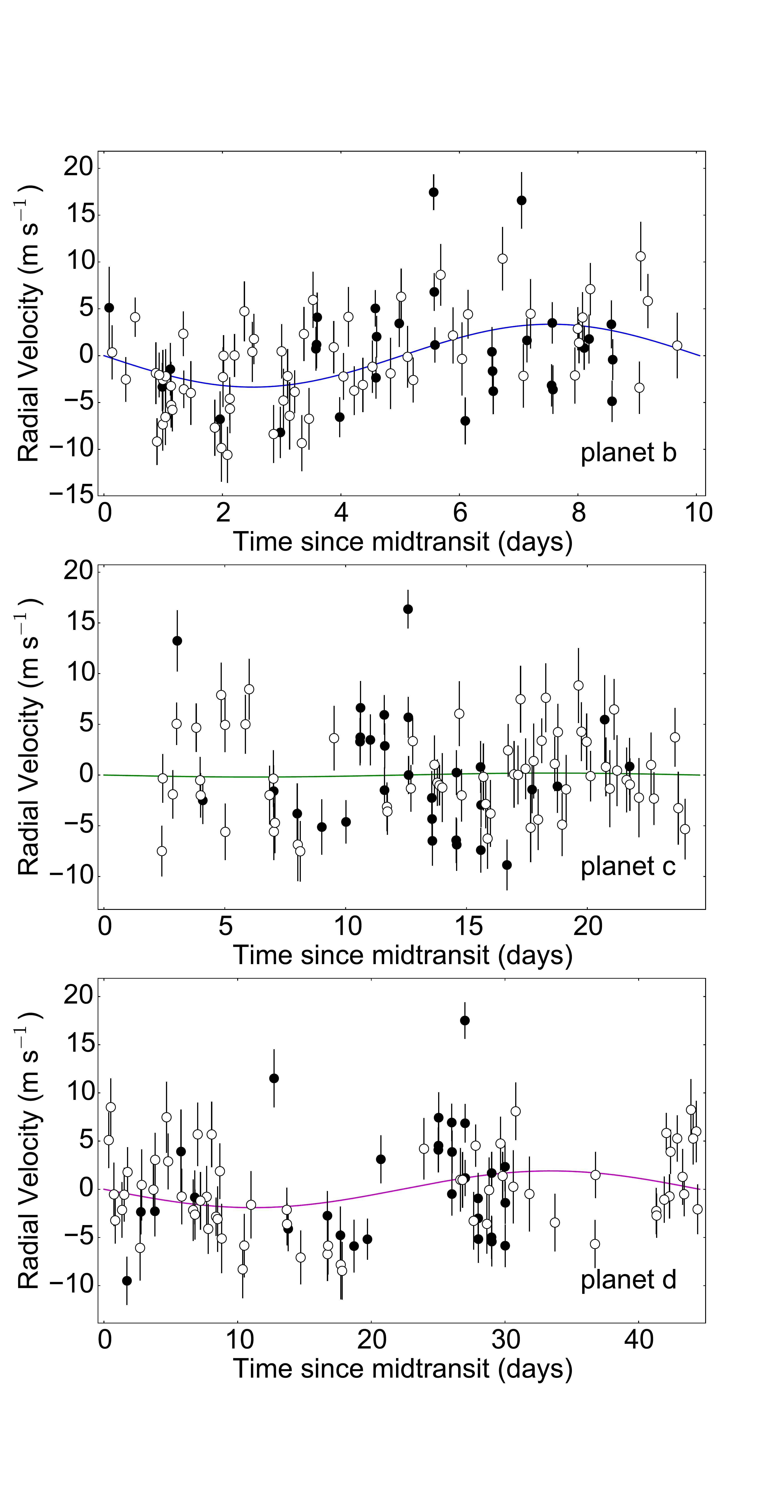}
\end{center}
\caption{Radial velocity as a function of time since mid-transit, for each of
  the planets in the K2-3 system (open circles are HARPS data from \citet{Almenara2015}; black circles are new PFS data). In each case, the modeled contributions of
  the other two planets has been removed, before plotting.}
\label{201367065fold}
\end{figure}

\begin{figure}
\begin{center}
\includegraphics[width= 0.99\columnwidth]{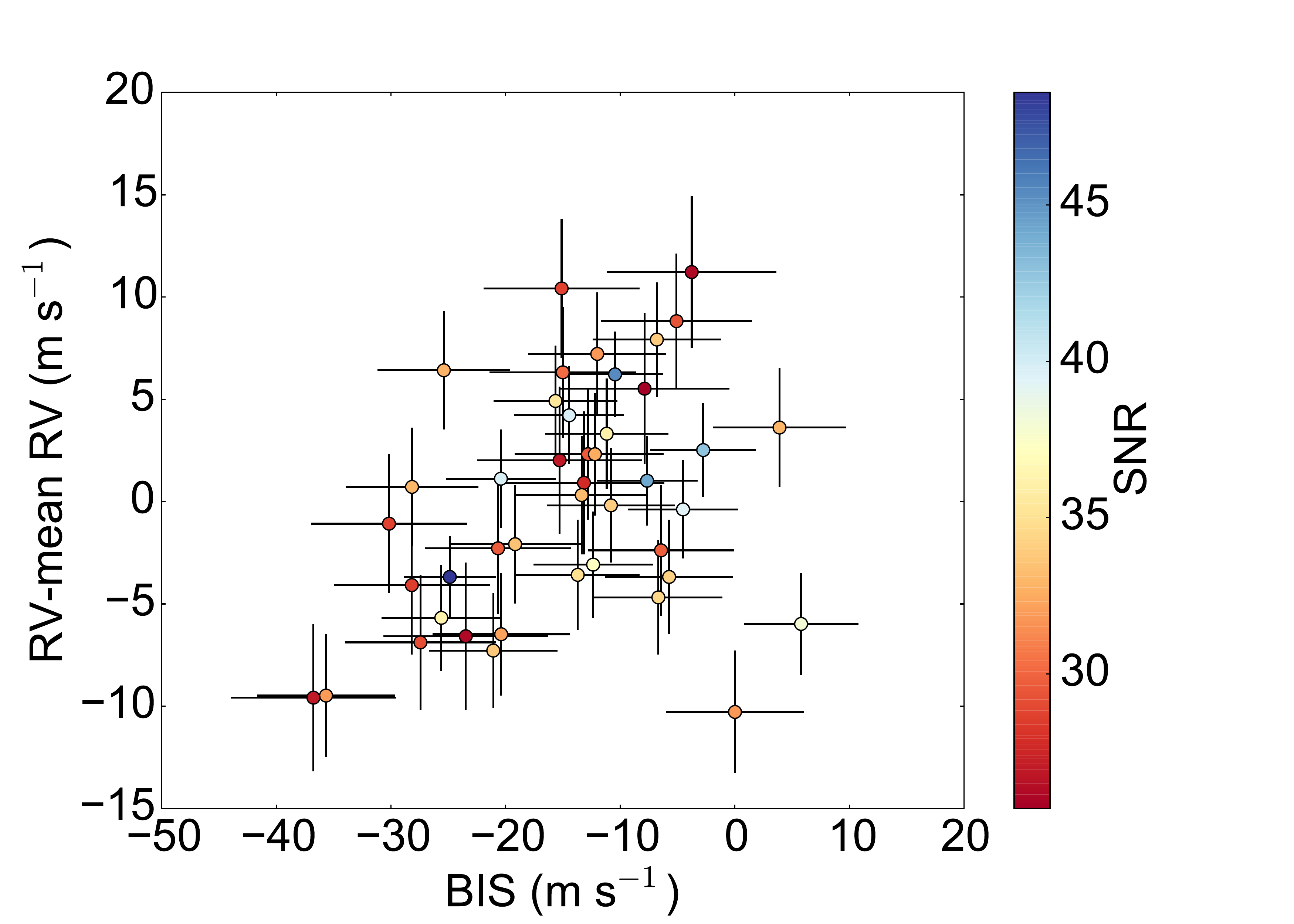}
\end{center}
\caption{The bisector span (BIS) and radial velocity variations from HARPS, for the K2-3 system. The color coding shows the SNR of the stellar spectra obtained near a wavelength range of 5500~Å. The uncertainty in the BIS was assumed to be twice the uncertainty in the radial velocity measurement. The Pearson correlation coefficient and the corresponding $p$-value were calculated to be 0.36 and 0.018. This suggests the presence of stellar activity signal, as the $p$-value indicates that the observed degree of correlation has only a 2\% chance of being produced by uncorrelated noise. See text for details.}
\label{201367065BIS}
\end{figure}

\begin{figure}
\begin{center}
\includegraphics[width= 1.1\columnwidth]{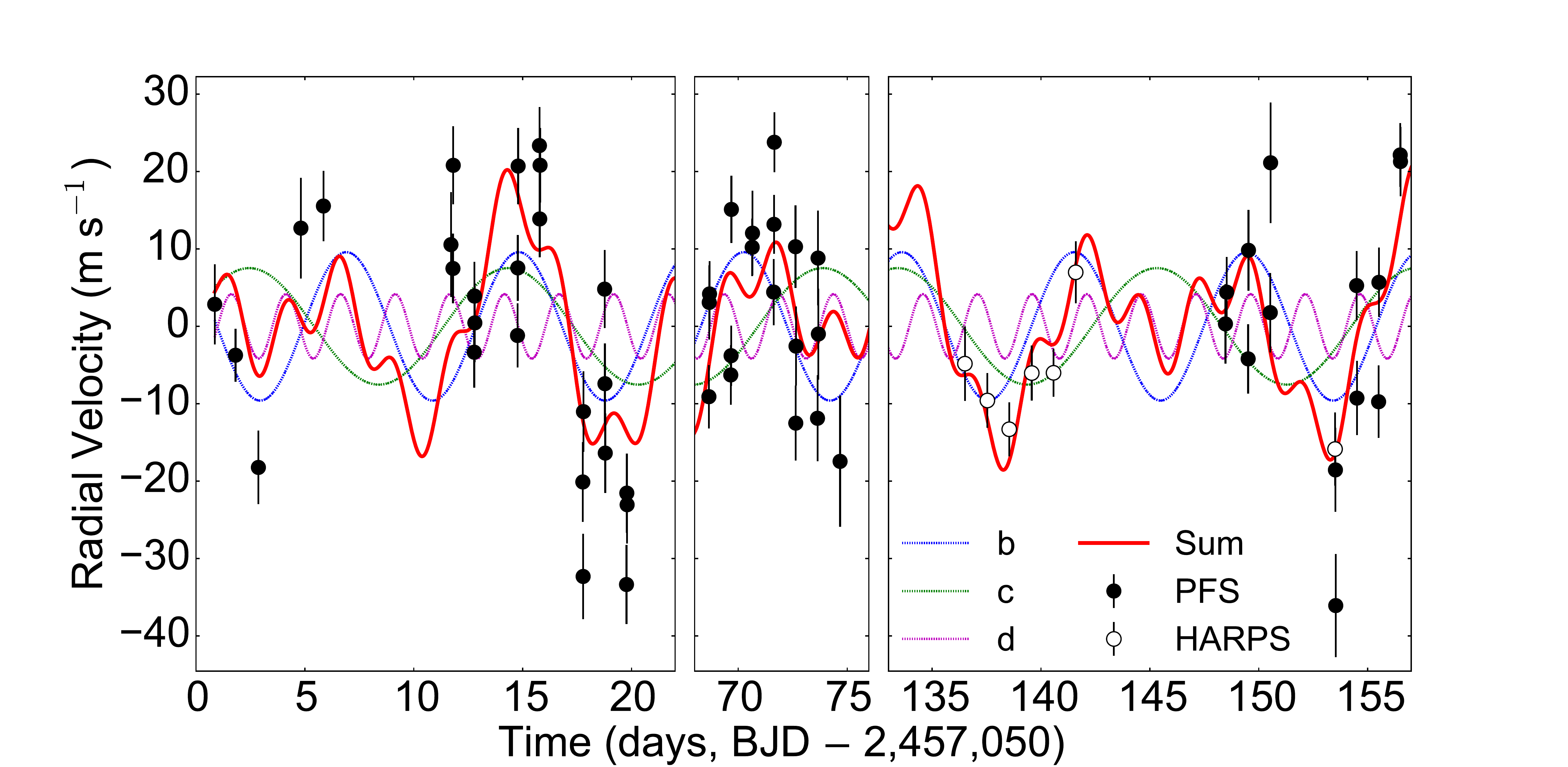}
\caption{Measured radial velocity of K2-19 (open circles are new HARPS data; black circles are new PFS data), and the best-fitting model (red line) assuming circular orbits.
The other colored lines show the contributions to the model curve from individual planets.
To account for radial velocity perturbations from stellar activity,
the data points within each 12-day interval were grouped together, and allowed to shift up or down by a constant velocity specific to the group (see text for details).}
\label{201505350_5offset}
\vspace{-0.1cm}
\end{center}
\end{figure}

\begin{figure}
\begin{center}
\includegraphics[width= 0.9\columnwidth]{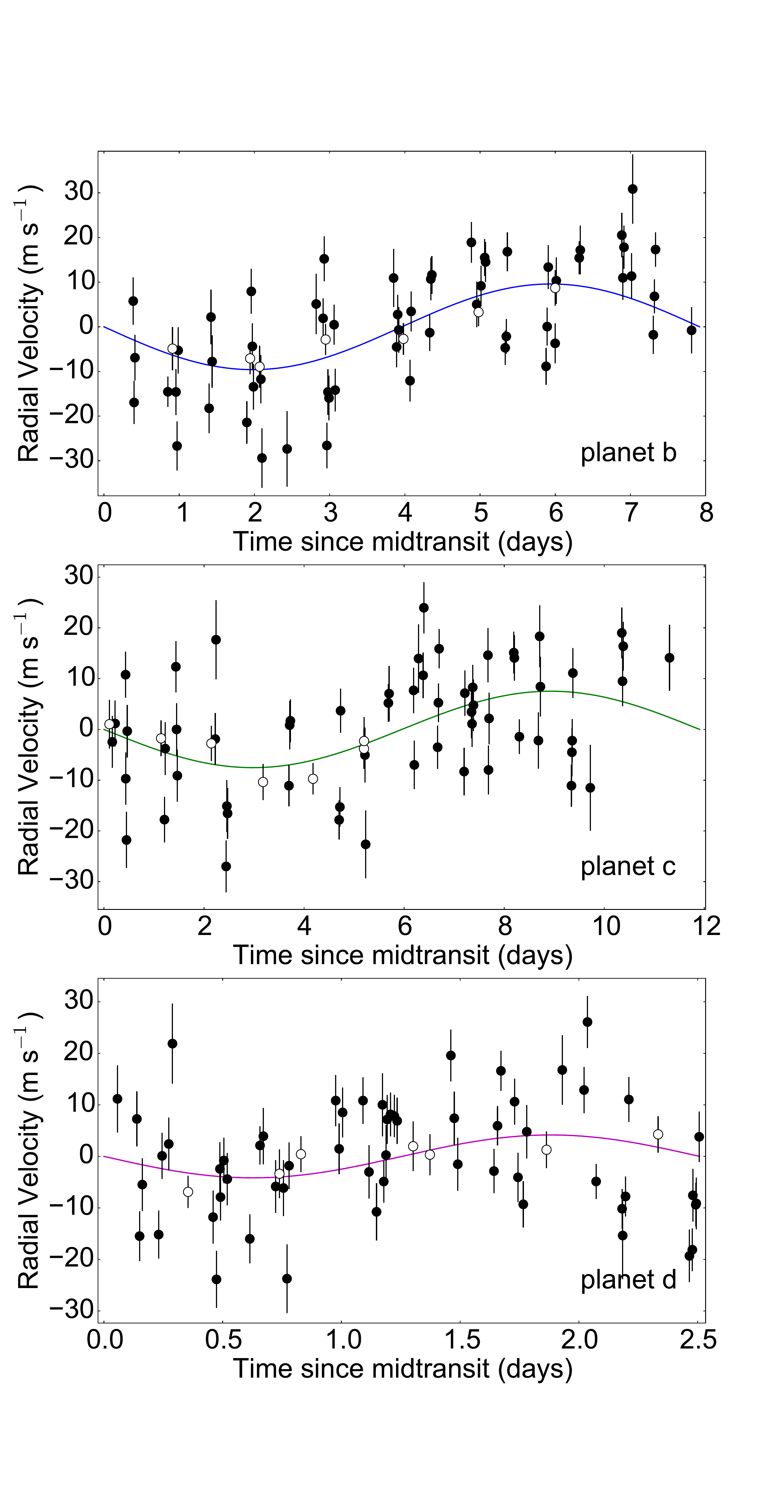}
\end{center}
\caption{Radial velocity as a function of time since mid-transit, for each of the planets in the K2-19 system (open circles are new HARPS data; black circles are new PFS data). For each planet, the modeled contributions of the other two planets has been removed, before plotting.}
\label{201505350fold}
\end{figure}

\begin{figure}
\begin{center}
\includegraphics[width= 0.99\columnwidth]{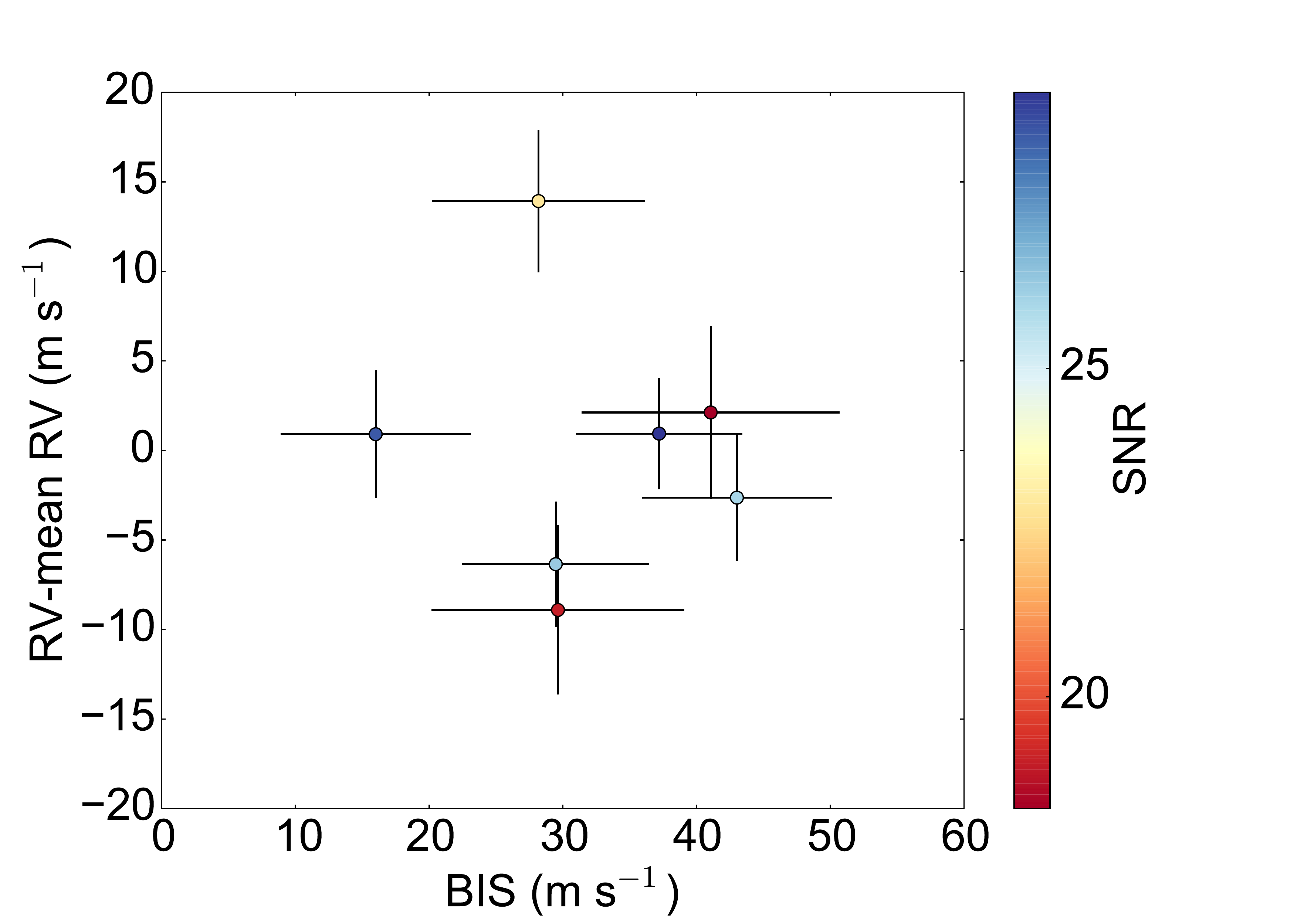}
\end{center}
\caption{The bisector span (BIS) and radial velocity variations from HARPS for the K2-19 system. The color coding shows the SNR of the stellar spectra obtained near a wavelength range of 5500~Å. The uncertainty in the BIS was assumed to be twice the uncertainty in the radial velocity measurement. No correlation was observed.}
\label{201505350BIS}
\end{figure}

\begin{figure}
\begin{center}
\includegraphics[width=1.1\columnwidth]{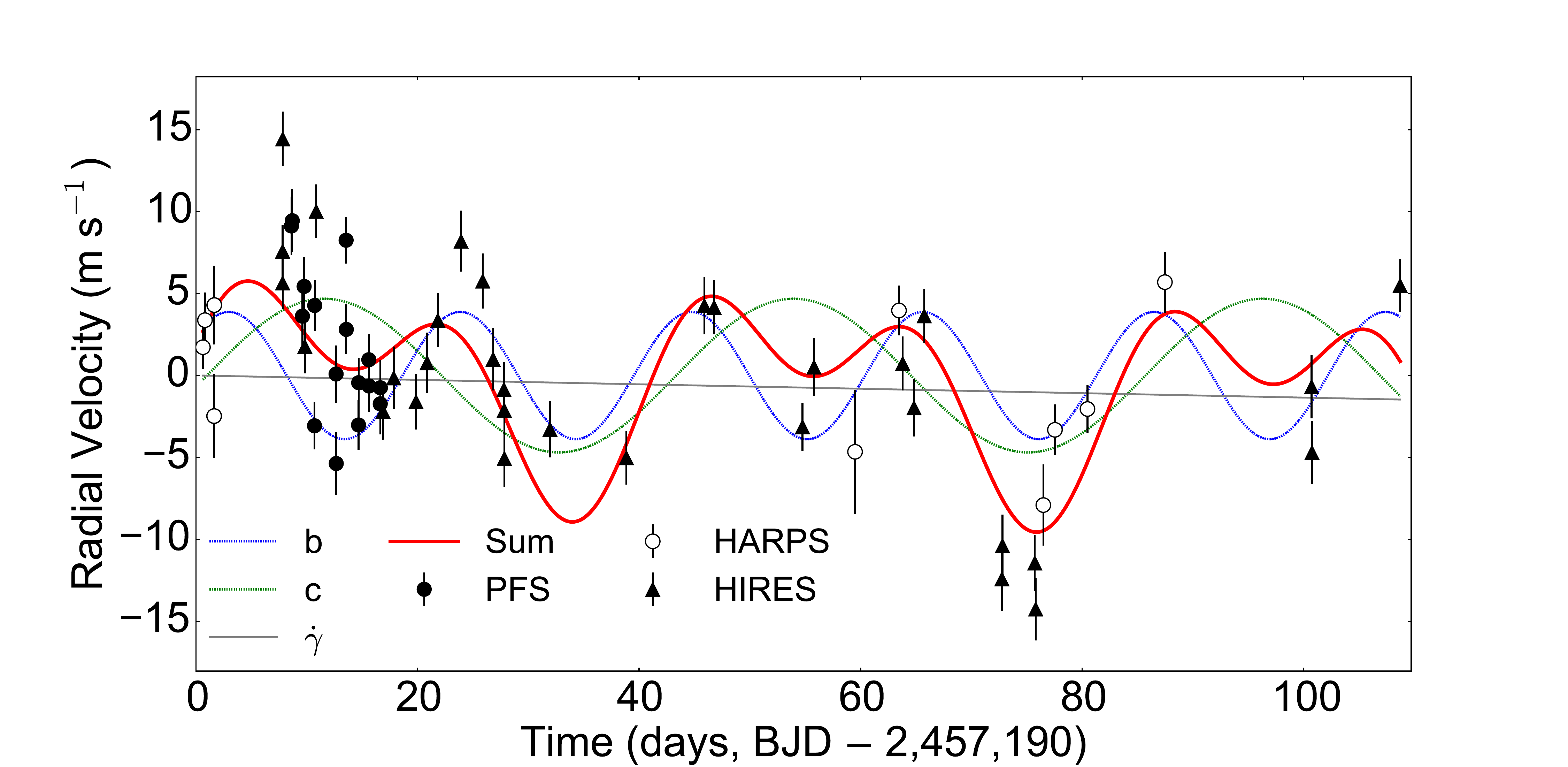}
\caption{Measured radial velocity of K2-24 (open circles are new HARPS data; black circles are new PFS data; triangles are HIRES data from \citet{Petigura2016}), and the best-fitting model
(red line) assuming circular orbits and an additional constant acceleration. The gray line shows the constant acceleration term. The other colored lines show the contributions to the model curve from individual planets.}
\label{203771098rvplot}

\vspace{-0.1cm}
\end{center}
\end{figure}

\begin{figure}
\begin{center}
\includegraphics[width=0.9 \columnwidth]{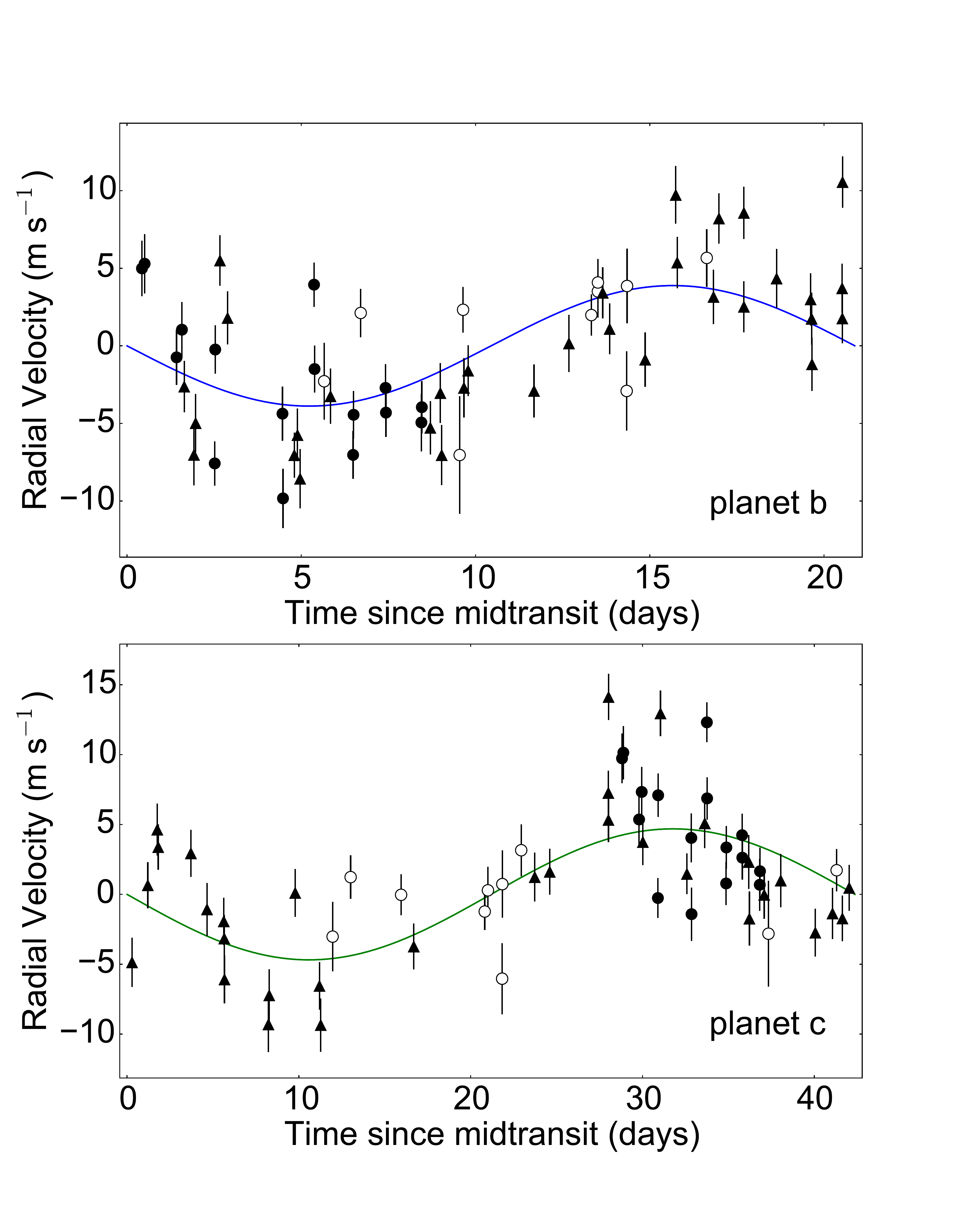}
\end{center}
\caption{Radial velocity as a function of time since mid-transit, for each planet in the K2-24 system (open circles are new HARPS data; black circles are new PFS data; triangles are HIRES data from \citet{Petigura2016}). In each case, the modeled contributions of
  the other planet has been removed, before plotting.}
\label{203771098fold}
\end{figure}

\begin{figure}
\begin{center}
\includegraphics[width= 0.99\columnwidth]{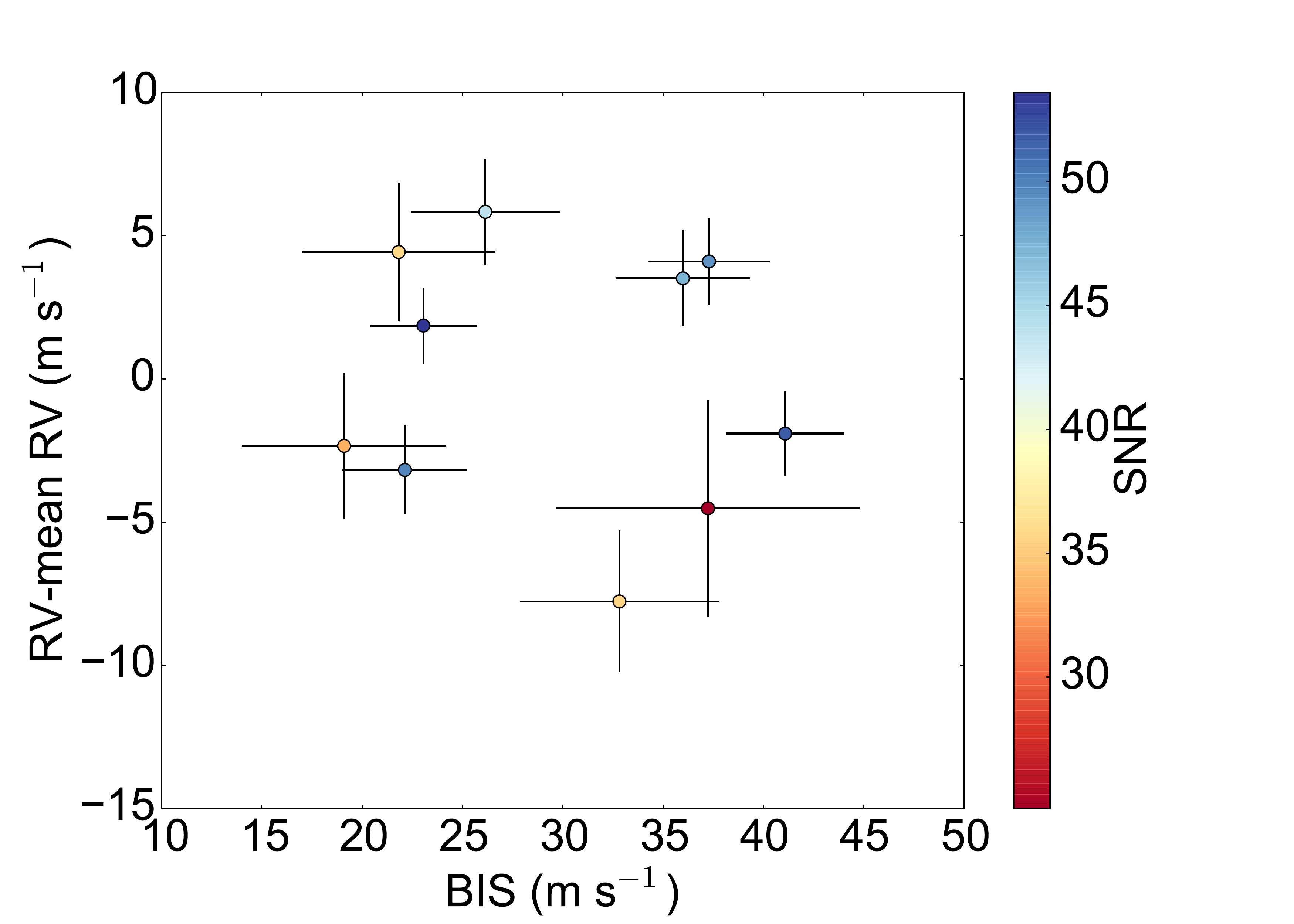}
\end{center}
\caption{The bisector span (BIS) and radial velocity variations from HARPS for the K2-24 system. The color coding shows the SNR of the stellar spectra obtained near a wavelength range of 5500~Å. The uncertainty in the BIS was assumed to be twice the uncertainty in the radial velocity measurement. The Pearson correlation coefficient and the corresponding p-value were calculated to be -0.14 and 0.70. The large p-value suggests no significant correlation between BIS and radial velocity variations.}
\label{203771098BIS}
\end{figure}

\section{Radial Velocity Analysis}

We modeled the Doppler measurements for each system as the sum of Keplerian radial velocity signals, one for each planet candidate identified in the {\it K2} photometry. We fixed the orbital periods and times of inferior conjunction at the values derived from {\it K2} photometry. We allowed the Doppler semi-amplitude $K$ induced by each planet to be a free parameter, with a uniform prior. We note in particular that we did not require $K$ to be positive. If we knew our model to be correct in all relevant details, and had a fundamental understanding of the uncertainties in the data points, the proper Bayesian approach would be to use a prior on $K$ that precludes negative values because they correspond to unphysical solutions (negative planet masses). However, given that our model may fail to include significant effects arising from stellar activity or additional planets, we opted not to place a prior on $K$ that would rule out negative values. In this way, when we find $K<0$ we will know the model is missing important sources of radial-velocity variation.

For each system we performed two fits: one in which the orbits were assumed to be circular, and one with eccentric orbits. In the first case, the radial velocity perturbation due to each planet is specified only by its Doppler semi-amplitude $K$. For eccentric models, the eccentricity $e$ and argument of periastron $\omega$ are also needed. To guard against the bias towards non-zero eccentricity \citep{Lucy1971}, we used the fitting parameters $\sqrt{e}\cos\omega$ and $\sqrt{e}\sin\omega$. For each observatory, we included a constant velocity offset $\gamma$ and a jitter parameter $\sigma_{\text{jit}}$ to subsume additional astrophysical and instrumental sources of apparent radial-velocity variation in excess of our internally-estimated measurement uncertainties. We also tested whether the inclusion of a constant acceleration term $\dot \gamma$ improved the model fit.

We adopted the following likelihood function:
\begin{equation}
\mathcal{L}=  \prod_{k=1}^{N_{\text{ins}}} \prod_{i=1}^{N_{\rm RV, k}}\left({\frac{1}{\sqrt{2 \pi (\sigma_i^2 + \sigma_{\text{jit}, ~k}^2)}} \exp \left[ - \frac{[RV(t_i) - \mathcal{M}(t_i)]^2}{2 (\sigma_i^2+\sigma_{\text{jit}, ~k}^2)}\right] }\right),
\end{equation}
where $RV(t_i)$ is the measured radial velocity
at time $t_i$; $\mathcal{M}(t_i)$ is the calculated radial velocity
at time $t_i$ for a particular choice of model parameters;
$\sigma_{i}$ is the internal measurement uncertainty;
 $\sigma_{\text{jit}, ~k}$ is the jitter for the $k$th instrument; $N_{\rm RV, k}$ is the number of observations obtained with the $k$th instrument and $N_{\text{ins}}$ is the number of instruments involved for each system. Uniform priors were adopted for all model parameters.

We maximized the likelihood using the Nelder-Mead (``Amoeba'') method as implemented in the Python {\tt scipy} package.  To determine the parameter uncertainies and covariances, we employed a Markov Chain Monte Carlo method. We used the affine-invariant ensemble sampler proposed by Goodman \& Weare (2010) and implemented in the Python package {\tt emcee} \citep{ForemanMackey2013}. We started 100 walkers in a Gaussian ball surrounding the best-fitting model parameters as obtained from the maximum likelihood estimation above. We stopped the walkers after 5000 links and checked the convergence by ensuring that the Gelman-Rubin potential scale reduction factor (Gelman et al.\ 1993) dropped below 1.03. We report the median of the marginalized posterior distribution, and defined the uncertainty interval based on the 16\% and 84\% percentile levels of the cumulative distribution.

We assessed the statistical significance of eccentric orbits and constant acceleration terms using the Bayesian Information Criterion, $BIC = -2\times \text{log}(\mathcal{L}_{\text{max}}) + N~\text{log}(M)$, where $\mathcal{L}_{\text{max}}$ is the maximum likelihood, $N$ is the number of parameters and $M$ is the number of observations \citep{Schwarz1978, Liddle2007}. As another measure of the significance of various models, we also report the root-mean-square (RMS) of the radial-velocity residuals for comparison with the RMS of the original radial velocity measurements.

To test whether the best-fitting model is likely to be dynamically stable, we calculated the orbital separations in units of mutual Hill radius:
\begin{equation}
\Delta = \frac{a_\mathrm{out}- a_\mathrm{in}}{R_H}.
\label{eqn:stability}
\end{equation}
\begin{equation}
R_H = \left[\frac{~M_\mathrm{out}+ ~M_\mathrm{in}}{3~M_\star}\right]^{1/3}\frac{\left(a_\mathrm{out}+ a_\mathrm{in}\right)}{2}.
\label{eqn:hill}
\end{equation}
where $M_\mathrm{in}$ and $M_\mathrm{out}$ are the masses of the inner and outer planets and $a_\mathrm{in}$ and $a_\mathrm{out}$ are their semi-major axes. For two-planet systems, we used the analytical criterion for stability:  $\Delta > 2\sqrt{3}$ (Gladman 1993). For systems with three planets, we used heuristic criterion for long-term dynamical stability proposed by \citet{Fabrycky2014}: $\Delta_1 + \Delta_2 > 18$ where $\Delta_1$ and  $\Delta_2$ are calculated respectively for the inner and outer pair of planets.

By modeling the observed Doppler shifts as the sum of Keplerian radial velocity signals, we implicitly assumed that gravitational interaction between planets can be ignored.  We tested and ultimately justified this assumption by experimenting with a fully dynamical ($N$-body) model obtained from the 4th order Hermite integration scheme that is available on the Systemic console \citep{Meschiari2009}. By maximizing the likelihood function with this dynamical model, we found that the best-fitting system parameters were consistent with those derived from our simpler multi-Keplerian model. We then examined the deviations between the radial velocities calculated in the dynamical model and the radial velocities in the Keplerian models. Over the timespan of our observations, the maximum deviation occurred for the K2-19 system and had a numerical value of 0.47~m~s$^{-1}$, which was much smaller than the uncertainties in the Doppler data ($\approx$3.8~m~s$^{-1}$). For all the other systems, the deviations were an order of magnitude smaller.
 
Some of the Doppler observations were conducted during an expected transit of one of the planets. Due to the scarcity and relatively large uncertainties of Doppler data, we did not attenmpt to model the Rossiter-McLaughlin (RM) effect in detail. Rather, we assessed the likely amplitude of the RM effect, and excluded from consideration the data for which the RM effect would make a significant contribution to the overall radial velocity.
We estimated the RM amplitude with the following equation:
 \begin{equation}
\Delta RV_{\rm{RM}} = \sqrt{1-b^2}\times\left(\frac{R_p}{R_{\star}}\right)^2\times v~\rm{sin}~i,
\label{eqn:RM}
\end{equation}
where $b$ is the impact parameter of the transit, $R_p/R_{\star}$ is the radius ratio between the planet and the host star, and $v~\rm{sin}~i$ is the projected rotational velocity of the stellar photosphere. For planets for which $\Delta RV_{\rm{RM}}$ was at least comparable to the measurement uncertainty, we chose to omit the data points taken during the calculated transit intervals.  The planets that were affected were K2-3b, K2-19b, EPIC~205071984b and EPIC~205071984c.
 
In order to test whether the observed Doppler signal could be associated with stellar activity, we plotted the measured Doppler shifts against the bisector span \citep[BIS, defined by][]{Queloz} derived from the cross-correlation functions of the obtained HARPS spectra. (Bisector data is not available for the other spectrographs.) We also calculated the Pearson correlation coefficient between the radial velocity and BIS, and the corresponding $p$-value which is roughly the probability that an uncorrelated process would produce a data set with a correlation coefficient at least as large as the observed coefficient. The BIS varies as a result of the deformation of stellar absorption lines induced by stellar activity. The gravitational pull of a planet should not change the BIS. Therefore, a strong correlation between BIS and Doppler shifts suggests that the observed Doppler shifts are the result of stellar activity rather than the gravitational influence of the planet.

\begin{figure}
\begin{center}
\includegraphics[width=1.1 \columnwidth]{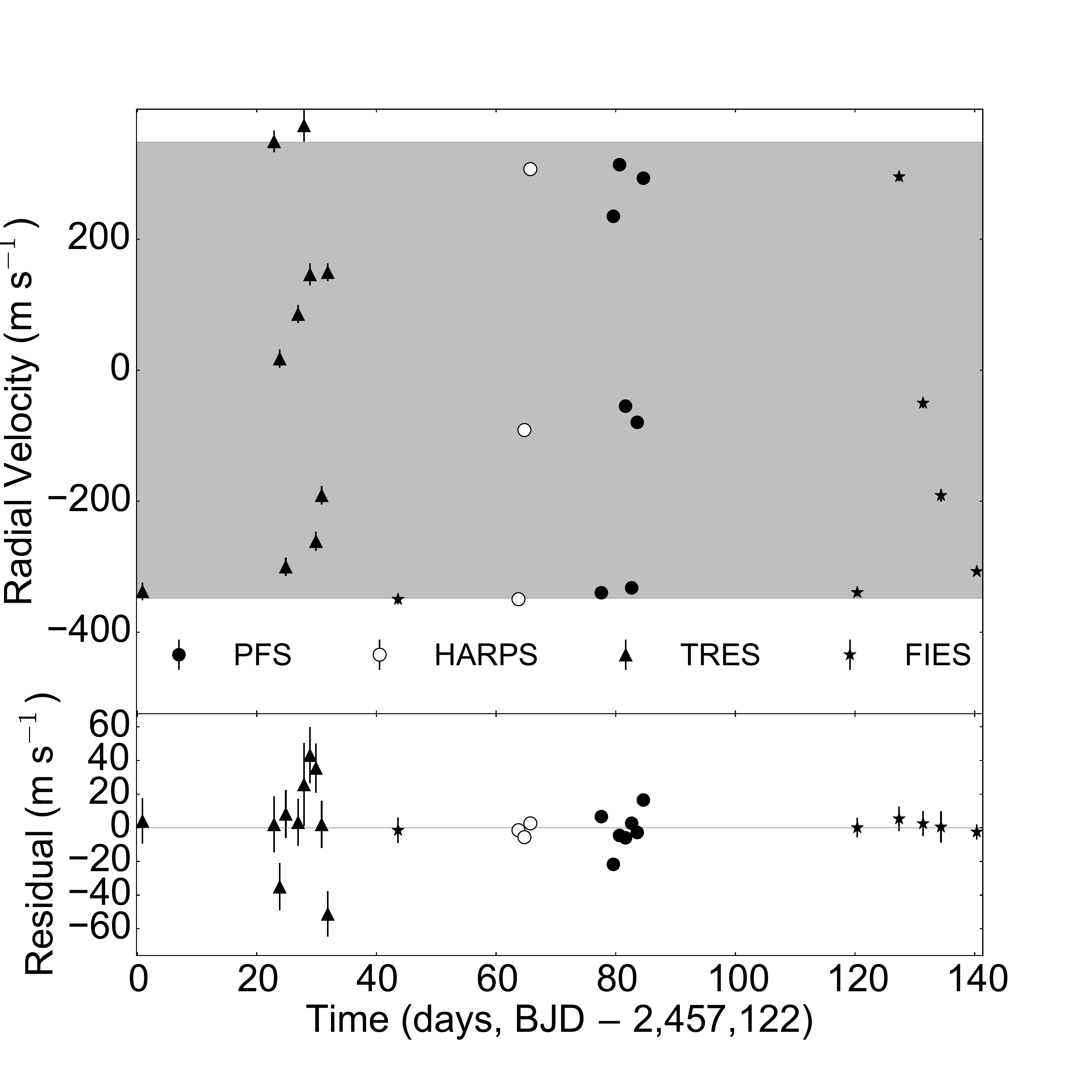}
\caption{{\it Top.}---Measured radial velocity of EPIC~204129699 (stars are FIES data from \citet{Grziwa2015}; open circles are HARPS data from \citet{Grziwa2015}; black circles are new PFS data; triangles are new TRES data). The gray rectangle illustrates the best-fitting $K$ value. (Because of the short orbital period, the model curve varies too rapidly to be plotted clearly over the entire time range.) {\it Bottom.}---The radial-velocity residuals, after subtracting the best-fitting model assuming a circular orbit.}
\label{204129699rvplot}
\vspace{-0.1cm}
\end{center}
\end{figure}

\begin{figure}
\begin{center}
\includegraphics[width=0.9 \columnwidth]{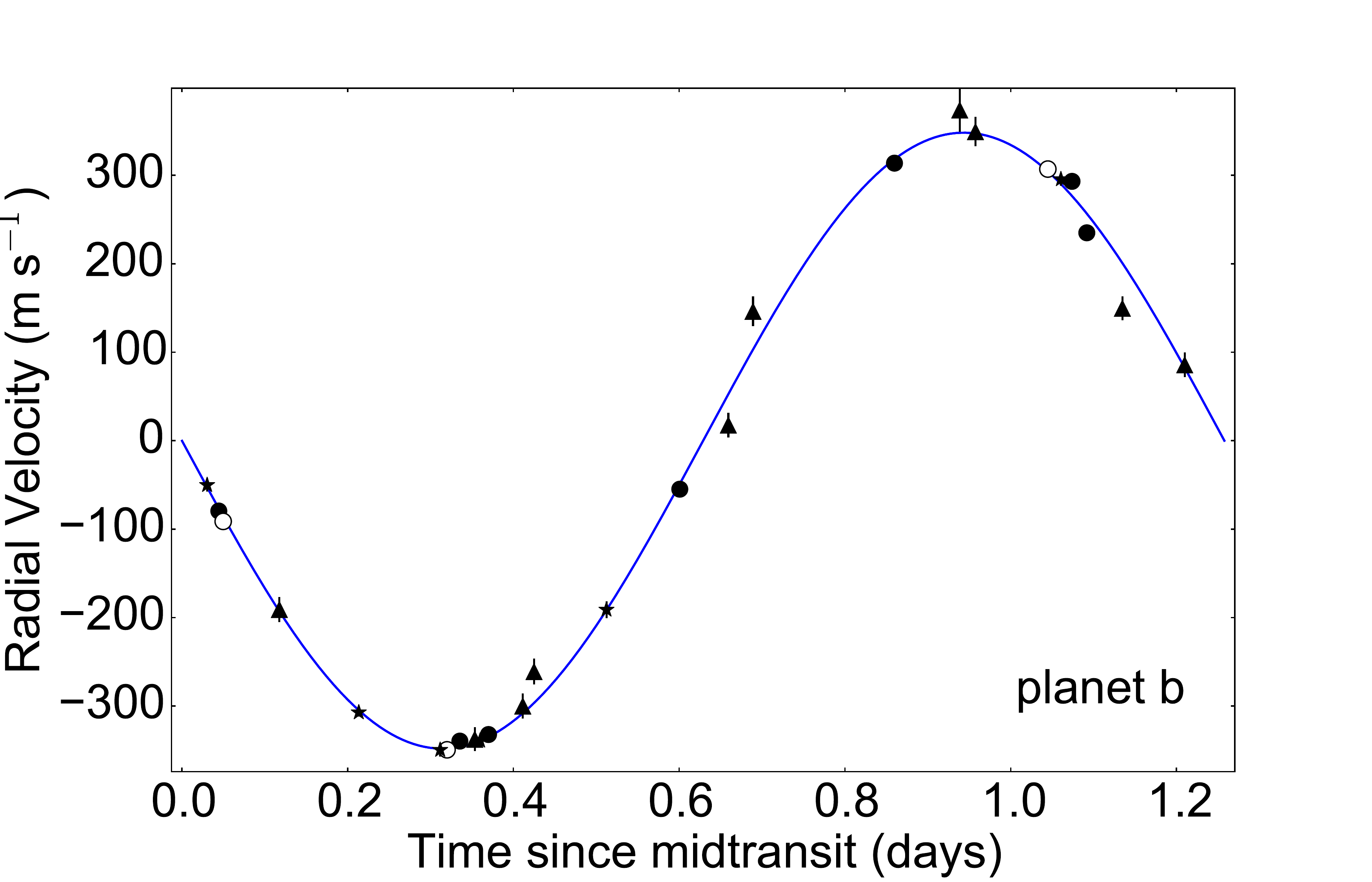}
\end{center}
\caption{Radial velocity of of EPIC~204129699 as a function of time since mid-transit of planet b (stars are FIES data from \citet{Grziwa2015}; open circles are HARPS data from \citet{Grziwa2015}; black circles are new PFS data; triangles are new TRES data).}
\label{204129699fold}
\end{figure}

\section{Individual Targets}

\subsection{K2-3}
K2-3 or EPIC~201367065 is a nearby M0V star with three transiting planets \citep{Crossfield2015}. The planets are on 10.1, 24.6 and 44.6 day orbits with radii of $2.18 \pm 0.30~R_{\oplus}$, $1.85 \pm 0.27~R_{\oplus}$  and $1.51 \pm 0.23~R_{\oplus}$. The outermost planet may lie within the habitable zone. The host star is relatively bright ($V = 12.17 \pm 0.01$) and small ($R_\star$ = $0.561 \pm 0.068 ~R_{\odot}$), thus making the system a favorable target for the James Webb Space Telescope ({\it JWST}). These planets have been validated by \citet{Crossfield2015}, \citet{Sinukoff2015} and \citet{Beichman} through adaptive optics imaging and Spitzer transit observations. 

We observed K2-3 with PFS from January 28th to April 11th, 2015. We gathered a total of 31 spectra. The typical internally-estimated measurement uncertainty was 2.5~m~s$^{-1}$. Initially
we modeled only the PFS data, and the results were not constraining.
For all three planets, we could only place upper bounds on the masses: $M_b$ < $15.1 ~M_{\oplus}$~(95\% conf.) $M_c$ < $6.3 ~M_{\oplus}$~(95\% conf.) $M_d$ < $21.1 ~M_{\oplus}$~(95\% conf.). Nevertheless, the PFS data did serve as an independent check of the measurements reported by \citet{Almenara2015}: $M_b$ = $8.4 \pm 2.1~M_{\oplus}$, $M_c$ = $2.1^{+2.1}_{-1.3}~M_{\oplus}$ and $M_d$ = $11.1^{+3.5}_{-3.5}~M_{\oplus}$.  Their results were based on the analysis of the 66 HARPS Doppler measurements with an average uncertainty of 2.9~m~s$^{-1}$. For all three planets, the PFS results were consistent with the HARPS results.  

We then performed a joint analysis using both the PFS and HARPS datasets. We started with the simplest model, assuming circular orbits. The radial-velocity residuals of the best-fitting model showed substantial temporal correlation. Specifically, the most prominent peak in the Lomb-Scargle periodogram of the radial velocity data occurred at $20.6 \pm 1.6$ days, with a false alarm probability (FAP) of $5.6 \times 10^{-3}$. This same peak remained the strongest in the periodogram of the radial-velocity residuals after subtracting the best-fitting model. Meanwhile, the stellar rotation period is $40 \pm 10$~days, based on the Lomb-Scargle periodogram of {\it K2} light curve. This suggests that the temporal correlation in the radial-velocity residuals might be caused by radial velocity perturbations induced by stellar activity. The BIS and the observed radial velocity variations from HARPS also showed substantial correlation (see Fig \ref{201367065BIS}). The Pearson correlation coefficient and the corresponding $p$-value were calculated to be 0.36 and 0.018, indicating that the observed degree of correlation has only a 2\% chance of being produced by uncorrelated noise.

To account for the effects of stellar variability, we tried the following strategy. We added to the model a series of sinusoidal functions of time with periods corresponding to the lowest harmonics of the stellar rotation period ($P_{\text{rot}}$, $P_{\text{rot}}/2$, $P_{\text{rot}}/3$, etc.). We imposed a prior on $P_{\text{rot}}$ ($40 \pm 10$~days). The amplitudes and phases of the sinusoids were allowed to float freely. In order to decide on the number of harmonics to include, we used the $BIC$ as a determinant. The model with the lowest $BIC$ was selected. After experimenting with different numbers of sinusoids to model the stellar activity signal, we found that none of the models with sinusoidal stellar activity signals yielded a lower $BIC$ number than the original model. In other words, the models with additional sinusoidal variability representing stellar activity did not improve the fit by enough to justify the increase in the number of parameters. Additionally, the orbital period of K2-3d and the stellar rotation period (45 and 40 days) are close to one another. Thus, the amplitude of the radial velocity signal from K2-3d and the amplitude of rotation-induced radial velocity perturbation are highly degenerate. A much longer time series would be required to disentangle the signals from K2-3d and stellar activity. Therefore, we reverted to the original model (see Fig.~\ref{201367065rvplot} and \ref{201367065fold}) and concluded that the mass of K2-3d cannot be constrained with the data at hand. The results for the other planets are $M_b$ = $8.1^{+2.0}_{-1.9}~M_{\oplus}$ and $M_c$ = $<4.2~M_{\oplus}$~(95\% ~conf.). The standard deviation of the radial velocities was $5.5$~m~s$^{-1}$.
After subtracting the best-fitting three-planet model, the standard deviation of the residuals was reduced to $4.7$~m~s$^{-1}$.

We then allowed for non-zero eccentricity. The mass constraints changed to: $M_b$ = $7.7^{+2.0}_{-2.0}~M_{\oplus}$, $M_c$ = $<12.6~M_{\oplus}$~(95\% ~conf.). See Table \ref{201367065para}. The eccentricity of planet b was constrained as $e_b = 0.21^{+0.15}_{-0.12}$, while the eccentricities of the planet c and d were unconstrained.
The circular model had a more favorable $BIC$ than the eccentric model. 

\begin{figure}
\begin{center}
\includegraphics[width=1.1\columnwidth]{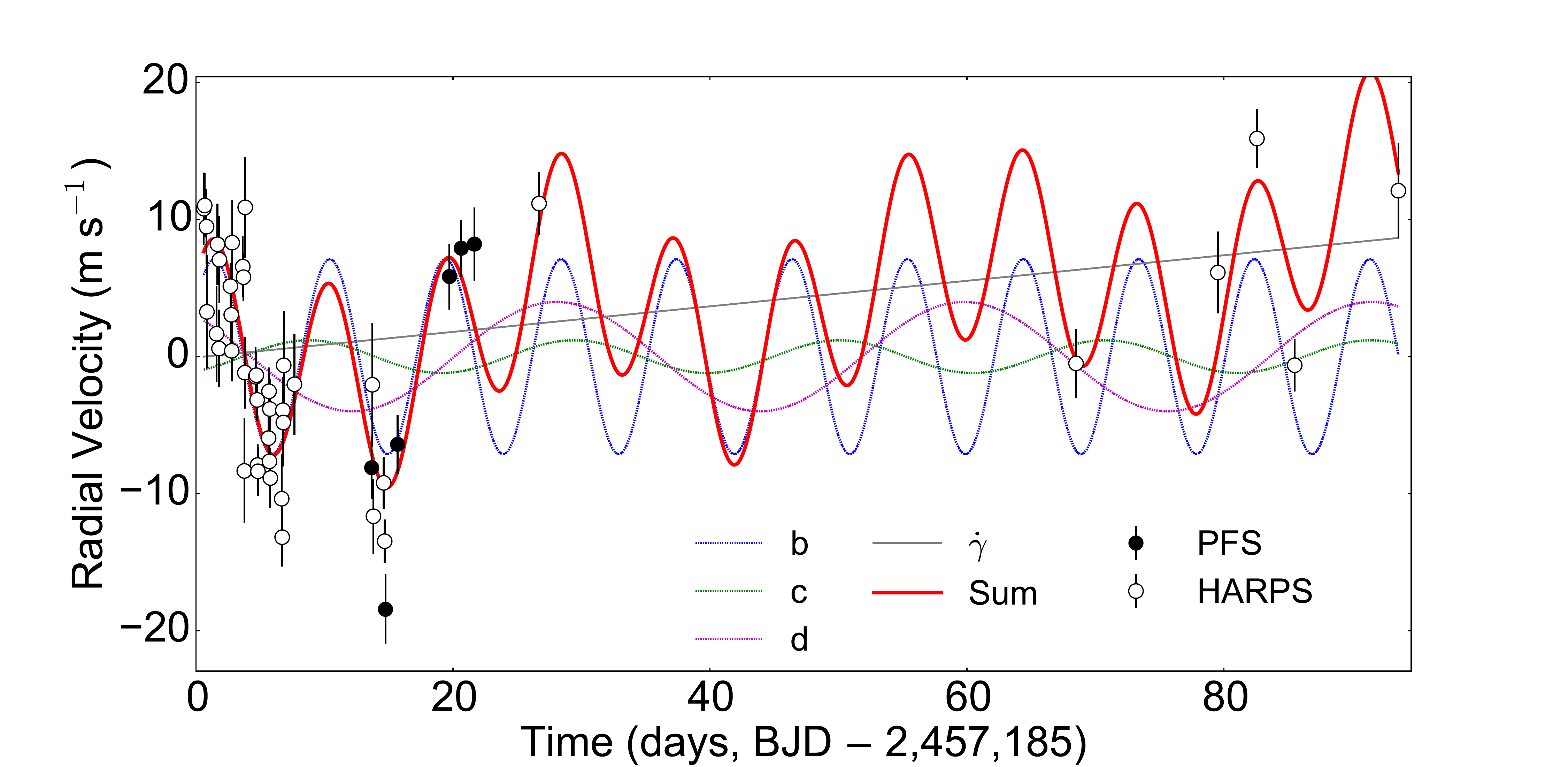}
\caption{Measured radial velocity of EPIC~205071984 (open circles are new HARPS data; black circles are new PFS data) and the best-fitting circular model (red line).  The contributions from each planet are also plotted as colored lines. The gray solid line represent a constant acceleration term.}
\label{205071984rvplot}
\vspace{-0.1cm}
\end{center}
\end{figure}

\begin{figure}
\begin{center}
\includegraphics[width=0.9\columnwidth]{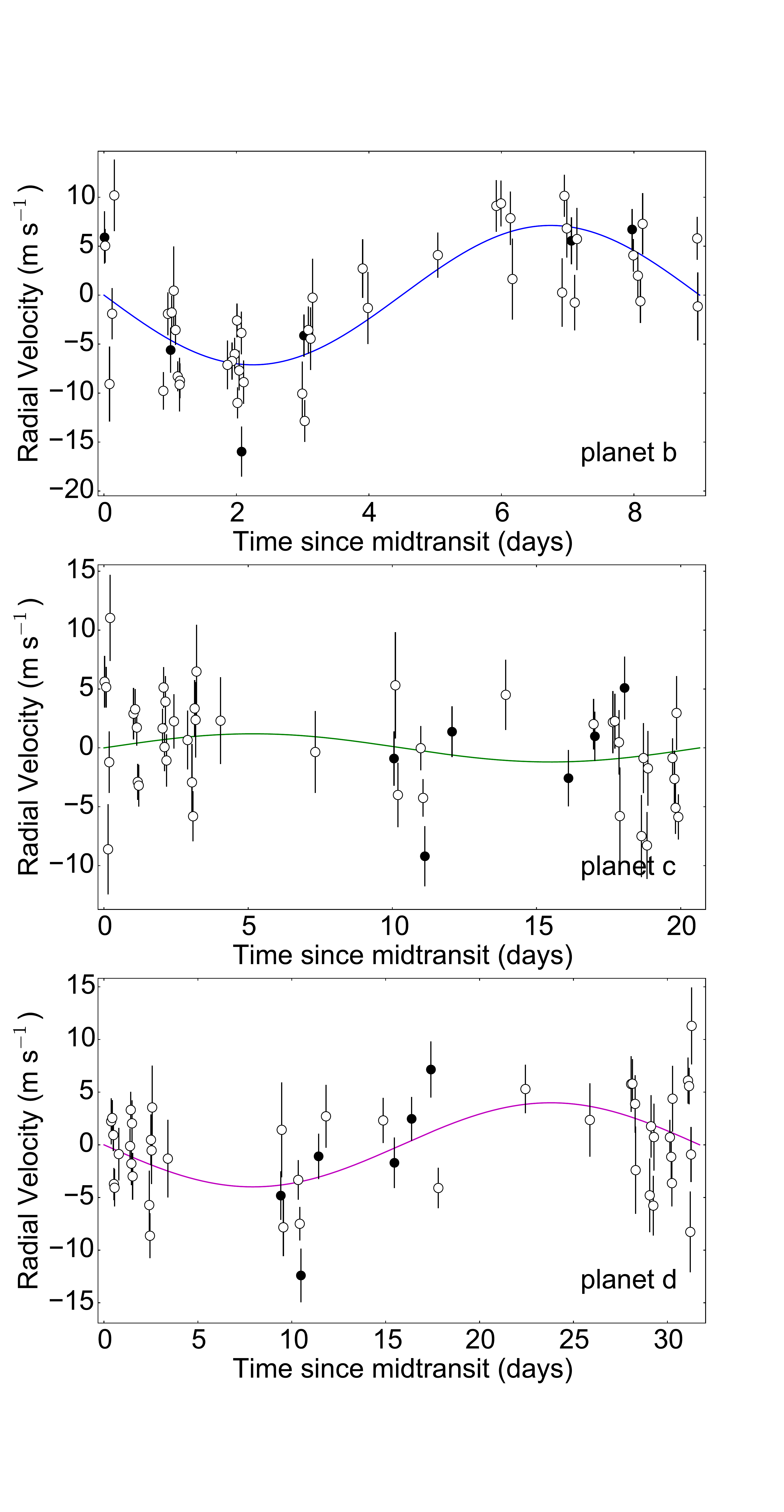}
\end{center}
\caption{Radial velocity as a function of time since mid-transit, for each of
the planets in the EPIC~205071984 system (open circles are new HARPS data; black circles are new PFS data). In each case, the modeled contributions of
  the other two planets has been removed, before plotting. In the best-fitting model, planet c has a negative mass, an unphysical result that probably arises from astrophysical or systematic noise.}
\label{205071984fold}
\end{figure}

\begin{figure}
\begin{center}
\includegraphics[width= 0.99\columnwidth]{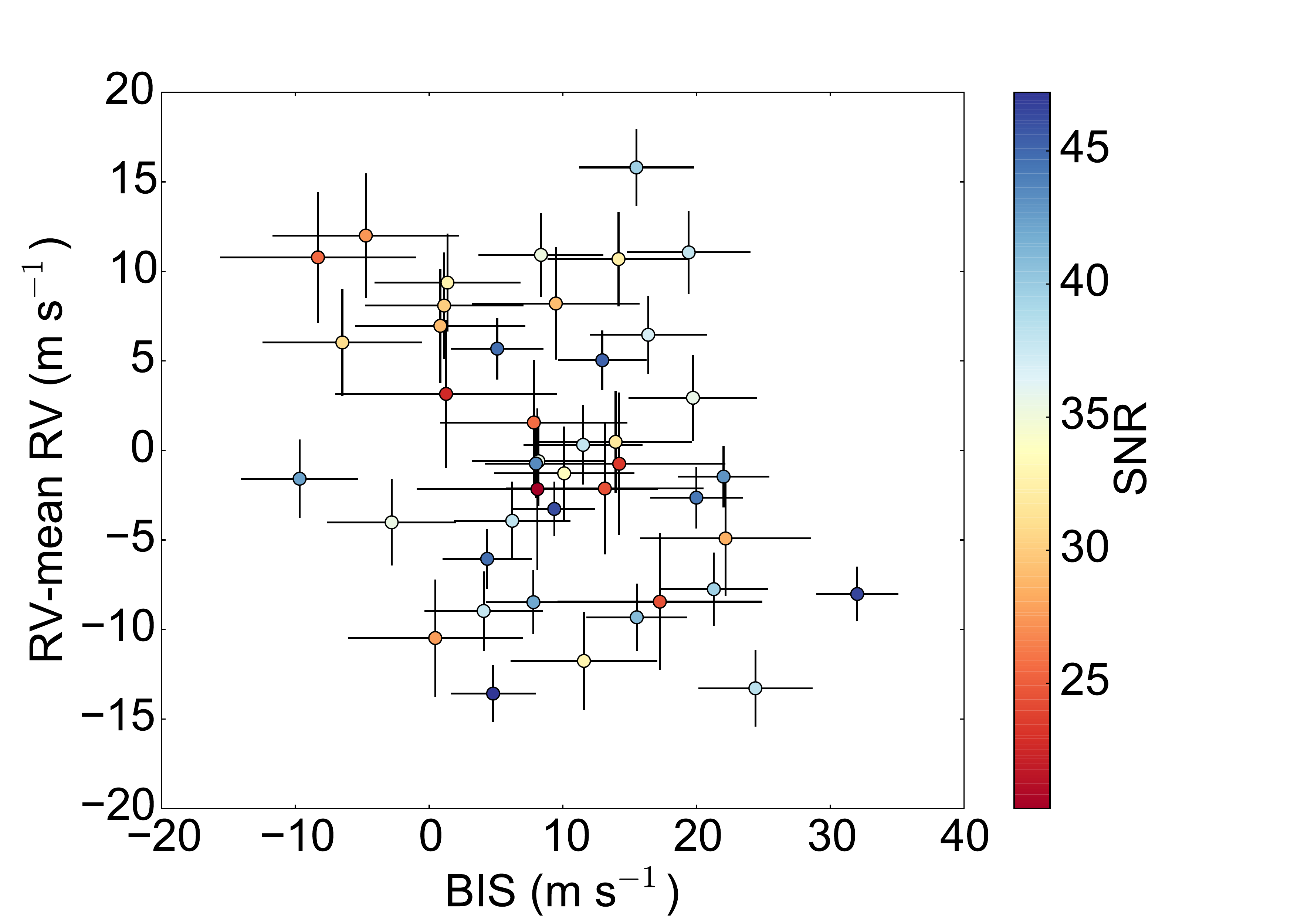}
\end{center}
\caption{The BIS and radial velocity variations from HARPS for the EPIC~205071984 system. The color coding shows the SNR of the stellar spectra obtained near a wavelength range of 5500~Å. The uncertainty in the BIS is taken as twice the uncertainty of the radial velocity measurement. The Pearson correlation coefficient between RV and BIS were calculated to be $-0.26$ with $p=0.094$.}
\label{205071984BIS}
\end{figure}

\subsection{K2-19}
K2-19 or EPIC~201505350 is a G9V star with three transiting planet candidates. The outer two planets (``b'', with radius $7.74 \pm 0.39~ R_{\oplus}$; and ``c'', $4.86^{+0.62}_{-0.44} ~R_{\oplus}$) have periods of 7.9~days and 11.9~days. They 
are within or near a 3:2 mean-motion resonance (MMR). These two planets were first reported by \citet{ForemanMackey2015}. \citet{Vanderburg2016} and \citet{Sinukoff2015} later revealed a third planet candidate in the system: ``d'', with a radius of $1.14 \pm 0.13 ~R_{\oplus}$ and an orbital period of 2.5~days. Note that the planet candidates were named based on the order in which they were discovered, rather than orbital distance. K2-19b and K2-19c have been validated using transit-timing variations and adaptive-optics imaging \citep{Armstrong2015,Narita2015, Sinukoff2015}. The transit candidate K2-19d has not been validated in this manner, but the false positive probability is expected to be low because it is associated with a star already known to have multiple transiting planets.

We observed K2-19 from January 28th to July 3rd, 2015, with PFS. We obtained a total of 61 spectra. The average internally-estimated radial velocity uncertainty was 5.0~m~s$^{-1}$. We also observed K2-19 with HARPS from June 12th to June 30th, 2015, obtaining 8 spectra. The average internally-estimated radial velocity uncertainty was 3.8~m~s$^{-1}$. Our radial velocity analysis started with the simplest model, assuming circular orbits for each planet candidate. The radial-velocity residuals showed strong temporal correlation, similar to the case of K2-3. Specifically, the most prominent peak in the Lomb-Scargle periodogram of the radial velocity data occurred at $21.6 \pm 1.1$~days with FAP~$= 1.7 \times 10^{-3}$. This is similar to the stellar rotation period of $20.4 \pm 2.7$~days, which was estimated based on the {\it K2} photometry. Thus, the $21.6 \pm 1.1$~day signal in the RV data seems likely to be caused by stellar activity. We did not observe a significant correlation between the measured RV and the BIS (see Fig \ref{201505350BIS}) but the test is not conclusive, given that only a few HARPS data points were obtained.

To account for the rotation-induced radial velocity perturbations, we experimented with the following strategies. We separated the Doppler observations into groups of 12-day duration and allowed each group to have an independent radial velocity offset. A similar grouping strategy was employed by \citet{Howard2013} to disentangle the planetary signal of Kepler-78b ($P_b\approx$~0.35~days) from the stellar activity signal ($P_{\text{rot}}\approx$~12.5 days). This grouping strategy is most effective when there is a clear separation in timescales between the planetary signal and the stellar activity signal. This separation is not as extreme for K2-19 as it was for Kepler-78b. The stellar rotation period of K2-19 is about 21 days, while the orbital periods of the planets are 2.5, 8 and 12 days. Nonetheless, the stellar rotation period of 21 days is still about twice as long as the longest orbital period of the planets. The choice of grouping in 12-day intervals was made such that each group covered at least one orbital period of the outermost planet. Therefore it should largely preserve the planetary signals. We also tried to model the stellar activity signal explicitly as a series of sinusoidal functions at the lowest harmonics of the stellar rotation period, but for our final analysis we adopted the grouping approach (see Fig \ref{201505350_5offset} and \ref{201505350fold}). This is because the grouping approach produced a better fit to the data as indicated by a $\Delta BIC$ of 43 for 68 data points. Additionally, the sinusoidal approach may not be appropriate for the case of K2-19, because the rotation-induced radial velocity perturbations might not remain coherent over the observation span of half a year ($\approx$8 rotation cycles).

The radial velocity signals of K2-19b and c were clearly detected with semi-amplitudes $K_b = 9.6^{+1.8}_{-1.6}$~m~s$^{-1}$ and $K_c = 7.5^{+2.1}_{-2.1}$~m~s$^{-1}$. Our analysis set an upper bound of $K_d < 6.9 $~m~s$^{-1}$~(95\%~conf.) on the semi-amplitude of the innermost and smallest planet K2-19d. The standard deviation of the measured radial velocities was $14.3$~m~s$^{-1}$, while the standard deviation of the residuals after subtracting the best-fitting model was reduced to $10.2$~m~s$^{-1}$. The planetary masses were then calculated from the semi-amplitudes, orbital periods, and the stellar mass. The masses of K2-19b-d were respectively constrained as $M_b = 28.5^{+5.4}_{-5.0} ~M_{\oplus}$, $M_c = 25.6^{+7.1}_{-7.1} ~M_{\oplus}$ and $M_d < 14.0 ~M_{\oplus}$~(95\%~conf.). We tested the long-term dynamical stability using Equations \ref{eqn:stability} and \ref{eqn:hill}. The best-fitting model has $\Delta_1 + \Delta_2 \approx 28>18$, consistent with stability.  For comparison, we also report the various results without applying the 12-day averaging method in Table \ref{201505350para}.

We also tried an eccentric model. The masses of planets b and c increased slightly while the constraint for planet d remained as an upper bound (see Table \ref{201505350para}). The eccentricities of planet b and c were both consistent with zero, with upper bounds of $e_b$ < $0.66$~(95\%~conf.) and $e_c$ < $0.82$~(95\% ~conf.). Obviously the eccentricity of planet d was not constrained, either.

\subsection{K2-24}
\citet{Petigura2016} reported the discovery of K2-24, a G9V star with two sub-Saturn planets close to (or within) a 2:1 MMR. The orbital periods are 21 and 42 days. They measured the planetary radii to be $R_b$ = $5.68 \pm 0.56 ~R_{\oplus}$ and $R_c$ = $7.82 \pm 0.72 ~R_{\oplus}$ by fitting transit models to the {\it K2} light curves. They were also able to validate the planets with adaptive-optics imaging and Doppler spectroscopy. They measured the masses, $M_b$ = $21.0 \pm 5.4 ~M_{\oplus}$ and $M_c$ = $27.0 \pm 6.9~M_{\oplus}$, based on Doppler data obtained with Keck/HIRES. They obtained 32 spectra with a typical internally-estimated radial velocity uncertainty of 1.7 m~s$^{-1}$. 

We observed K2-24 with PFS from June 25th to July 3rd, 2015, obtaining a total of 16 spectra. The average internally-estimated radial velocity uncertainty was 1.7 m~s$^{-1}$. We also monitored K2-24 with HARPS from June 17th to September 11th, 2015, obtaining 10 spectra. The average internally-estimated radial velocity uncertainty was 2.1 m~s$^{-1}$. We initially modeled the PFS and HARPS data only, as an independent check of the results reported by \citet{Petigura2016}. This led to mass constraints of $M_b$ = $30.0^{+9.1}_{-9.3} ~M_{\oplus}$ and $M_c$ < $21.3 ~M_{\oplus} $~(95\%~conf.). Both of these are consistent with the HIRES results. We then proceeded to perform a joint analysis with the HIRES data, yielding mass constraints of $M_b$ = $19.8^{+4.5}_{-4.4} ~M_{\oplus}$ and $M_c$ = $26.0^{+5.8}_{-6.1} ~M_{\oplus}$ (see Fig.~\ref{203771098rvplot} and \ref{203771098fold}). \citet{Petigura2016} found it necessary to include a constant acceleration term, $\dot \gamma = -22.5 \pm 9.2 $~m~s$^{-1}$yr$^{-1}$, to obtain a satisfactory fit. This may indicate the presence of an additional companion in the system. Our joint analysis led to reduced significance of this acceleration: $\dot \gamma = -12^{+10}_{-10}$~m~s$^{-1}$yr$^{-1}$. In particular, the HARPS data which spanned $\approx$3~months seem to disfavor a constant acceleration (see Fig.~\ref{203771098rvplot}). When we allowed for non-zero eccentricity for both planets, the best-fitting planet masses increased slightly (see Table~\ref{203771098para}). The eccentricity constraint for the inner planet's orbit was $0.24^{+0.10}_{-0.11} $ \citep[$0.24 ^{+0.11}_{-0.11}$]{Petigura2016}. The eccentricity of the the outer planet's orbit was consistent with zero with an upper bound of $ < 0.58 $~(95\%~conf.)  \citep[$ < 0.39 $~(95\%~conf.)]{Petigura2016}. We adopted the circular models as they were favored by a $\Delta BIC$ of 14.

We checked if the observed radial velocity could be caused by stellar activity. The Pearson correlation coefficient and the corresponding $p$-value were calculated to be $-0.14$ and 0.70. The large $p$-value suggests that there is no significant correlation between BIS and radial-velocity variations (see Fig \ref{203771098BIS}). The standard deviation of the radial velocity data was 6.2~m~s$^{-1}$, while the standard deviation of the residuals after subtracting the best-fitting circular model reduced to 3.2~m~s$^{-1}$. The separation between the two planets, in units of their mutual Hill radius, is $\Delta \approx $ 16.9 (as defined in Equations \ref{eqn:stability} and \ref{eqn:hill}). This is well above the minimum value of $2\sqrt3$ that is needed for dynamical stability, according to the criterion of Gladman~(1993).

\subsection{EPIC~204129699}
EPIC~204129699 is G7V star with a 1.26-day hot Jupiter. \citet{Grziwa2015} modeled the {\it K2} photometry and found that the transit trajectory of the planet
likely grazes the stellar limb ($b =0.9-1.05$).
They also measured the planetary mass to be $M_b = 1.774 \pm 0.079~M_{\text{Jup}}$ using Doppler observations obtained with FIES and HARPS. We obtained 7 PFS spectra from June 26th to July 3rd, 2015, and 10 TRES spectra from April 10th to May 11th, 2015. The average internally-estimated radial velocity uncertainties for PFS and TRES were 1.5~m~s$^{-1}$ and 16~m~s$^{-1}$, respectively.

To improve on our knowledge of the stellar parameters for EPIC 204129699, we also obtained a high-SNR spectrum with the High Dispersion Spectrograph (HDS) on the Subaru 8.2m telescope on May 30, 2015. From the measurement of equivalent widths of iron lines \citep{2002PASJ...54..451T}, we estimated the stellar atmospheric parameters as $T_\mathrm{eff}=5384\pm 46$ K, $\log g=4.410\pm0.060$, and $\mathrm{[Fe/H]}=0.21\pm0.03$.  We also analyzed the 10 TRES spectra using the procedure described by \citet{Buchhave2012} and \citet{Buchhave2015}. The results ($T_\mathrm{eff}=5445\pm49$ K, $\log g=4.52\pm0.10$, and $\mathrm{[Fe/H]}=0.16\pm0.08$) were consistent with the HDS values. We thus opted to combine these measurements to obtain a weighted average of $T_\mathrm{eff}=5412\pm33$ K, $\log g=4.44\pm0.05$, and $\mathrm{[Fe/H]}=0.20\pm0.03$. These atmospheric parameters were then converted into estimates for the stellar mass and radius using the empirical relations presented by \citet{2010A&ARv..18...67T}. Using these relations and Monte-Carlo sampling \citep{2012ApJ...756...66H} we obtained $M_\star=1.000\pm0.064M_\odot$ and $R_\star=0.986\pm0.070R_\odot$. These stellar parameters are consistent with the Yonsei-Yale stellar-evolutionary models \citep{2001ApJS..136..417Y}. 

By analyzing our PFS and TRES data, we measured the mass of EPIC~204129699b to be $M_b = 1.856 \pm 0.084 ~M_{\text{Jup}}$. This is consistent with the results of \citet{Grziwa2015}. We then analyzed our PFS and TRES data jointly with the published FIES and HARPS data from \citet{Grziwa2015}. The result for the planet mass is $M_b = 1.857 \pm 0.081~M_{\text{Jup}}$. The RMS of the measured radial velocities was 262~m~s$^{-1}$, whereas the RMS of residual radial velocities was reduced to 18.2~m~s$^{-1}$. Given the short tidal circularization timescale that is expected for such a close-in orbit, the assumption of a circular orbit seems justified. Indeed, our joint analysis favored the circular model over the eccentric model by $\Delta BIC$ of 15 for 25 data points, and placed an upper limit on the eccentricity as $<0.027$~(95\%~conf.). See Fig.~\ref{204129699rvplot} and \ref{204129699fold} and Table \ref{204129699para}.

\subsection{EPIC~205071984}
EPIC~205071984 is a G9V star with three transiting planet candidates reported by \citet{Vanderburg2016} and \citet{Sinukoff2015}. The inner planet is on a 9-day orbit with a radius of $5.38 \pm 0.35~R_{\oplus}$. The outer two planets have radii of $3.48 \pm 0.97 ~R_{\oplus}$ and $3.75 \pm 0.40 ~R_{\oplus}$ and they are within or close to a 3:2 MMR, with orbital periods of 20.7 and 31.7~days. Adaptive-optics images obtained by \citet{Sinukoff2015} revealed several nearby faint sources; however, through a careful analysis of the {\it K2} pixel data, those authors demonstrated that the dimming events cannot be associated with those faint sources, and are most likely associated with EPIC~205071984.

We monitored EPIC~205071984 with PFS from June 25th to July 3rd, 2015, obtaining 6 spectra. The average internally-estimated radial velocity uncertainty was 2.4 m~s$^{-1}$. We also observed EPIC~205071984 with HARPS from April 5th to July 7th, 2015, obtaining 43 spectra. The average internally-estimated radial velocity uncertainty was 2.6 m~s$^{-1}$. Our analysis started with the simplest model assuming circular orbits. The residuals suggested a positive linear trend with time. We added a term to the model representing a constant acceleration, which improved the model fit by $\Delta BIC~\approx 12$ (see Figs.\ref{205071984rvplot} and \ref{205071984fold}). The resulting mass for planet b was $M_b = 21.1^{+5.9}_{-5.9}~M_{\oplus}$ corresponding to a 3.5$\sigma$ detection. We did not detect planets c and d.  In fact, the best-fitting model suggested an unphysical negative mass for planet c (see Fig.~\ref{205071984fold}). We therefore only report upper limits for planet c and d: $M_c$ < $8.1 ~M_{\oplus}$ and $M_d$ < $35.0 ~M_{\oplus}$ both at a 95\% confidence level. The RMS of the measured radial velocities was 7.9~m~s$^{-1}$, whereas the RMS of residual radial velocities was reduced to 4.2~m~s$^{-1}$. The Pearson correlation coefficient between RV and BIS is $-0.26$ with $p=0.094$ (see Fig.~\ref{205071984BIS}).

The constant acceleration term was constrained as $\dot \gamma$ = $34.0^{+9.9}_{-9.7} $~m~s$^{-1}$yr$^{-1}$. If this trend is associated with another companion in the system, we can estimate the order of magnitude of $M~\text{sin}~i$ as function of semi-major axis $a$. Assuming a circular orbit and $M~\text{sin}~i \ll ~M_\star$, then $\dot \gamma \approx G~ M~\text{sin}~i /a^2$. The data suggest that the companion's mass and semi-major axis are such that
\begin{equation}
M~\text{sin}~i \approx 60~M_{\oplus} \left(\frac{a}{1~{\rm AU}}\right)^2.
\end{equation}

We then tried allowing for eccentric orbits for the three planets. The mass of planet b increased to  $25.0^{+7.4}_{-7.0} ~M_{\oplus}  $, while the upper bounds on planet c and d also increased slightly (see Table \ref{205071984para}). We found that the eccentricity of planet b was consistent with zero with an upper bound of $<$0.43 (95\%~conf.). Due to non-detection of the signals from planet c and d, their eccentricities were unconstrained.  

\section{Discussion}
\begin{figure}
\begin{center}
\includegraphics[width=1.1 \columnwidth]{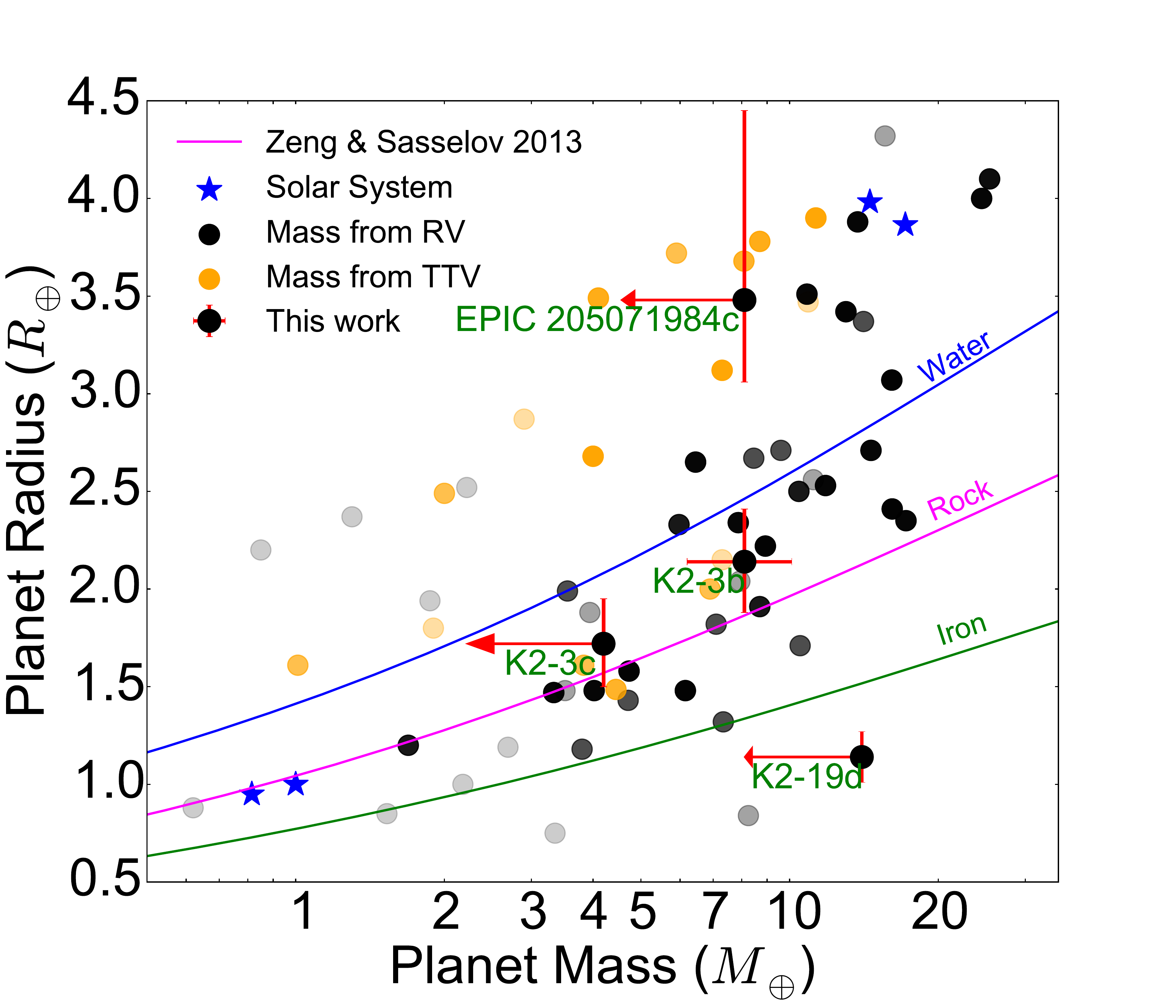}
\end{center}
\caption{The mass radius diagram of sub-Neptune planets. The yellow circles indicate the mass measurements from TTV analysis; the black circles indicate the mass measurements from Doppler method. The opacity indicates the quadrature sum of the SNR of the mass and radius measurement of each planet. The mass measurements and upper limits ($95\%$~conf.) presented in this paper are annotated with error bars and red arrows. The theoretical mass-radius relationships from \citet{Zeng2013} are also plotted. ``Rock'' corresponds to a composition of 100\% MgSiO$_3$.}
\label{rm}
\end{figure}

Fig.~\ref{rm} shows all of the reported measurements of mass and radius for planets smaller than $4 R_{\oplus}$, from our own sample and from the literature.  To convey the statistical significance of the measurements, we have set the opacity of the data points in proportion to the quadrature sum of the SNR of the radius and mass measurements for each planet. Darker symbols correspond to higher SNR.  We have also plotted theoretical mass-radius relationships for various compositions, from \citet{Zeng2013}.

K2-3 is a favorable target for {\it JWST} owing to the fact that the host star is relatively bright ($K_s = 8.561 \pm 0.023$) and small ($R_\star$ = $0.561 \pm 0.068 ~R_{\odot}$). Our analysis has confirmed the mass measurement of the innermost planet K2-3b by \citet{Almenara2015}: they found $M_b$ = $8.4 \pm 2.1~M_{\oplus}$  while we find $M_b$ = $8.1^{+2.0}_{-1.9}~M_{\oplus}$.  Because we adopted a slightly different stellar mass from that reported by \citet{Almenara2015}, a more direct comparison is between the measured velocity semi-amplitudes $K$.  \citet{Almenara2015} reported $K_b$ = $3.60 \pm 0.87$~m~s$^{-1}$, and our result of $K_b$ = $3.36^{+0.75}_{-0.74}$~m~s$^{-1}$ is consistent with theirs. The mass and radius of K2-3b are reminiscent of 55~Cnc~e. Both planets have dimensions of roughly $\approx$8~$M_{\oplus}$ and $\approx$2~$R_{\oplus}$. 

Using the model of \citet{Zeng2015}, the measured mass and radius of K2-3b suggests that the planet may contain up to 60\%~$\mathrm{H_2O}$ with 40\%~$\mathrm{MgSiO_3}$. Should the presence of $\mathrm{H_2O}$ be confirmed by spectroscopic follow-up, interesting questions will be raised. How did the planet acquire its water? Did it form beyond the snow line and migrated inward? Or did it form {\it in situ} and acquire its water from cometary impacts? For K2-3c, our analysis placed a upper bound of $<$$4.2~M_{\oplus}$ with 95\% confidence. This upper bound is not strong enough to allow for firm conclusions about the composition, but it is at least suggestive of a composition similar to or less dense than pure rock. As we noted in the previous section, the proximity of the orbital period of K2-3d and the stellar rotation period (45 days and 40 days) makes it difficult to disentangle the planetary signal of K2-3d and the stellar activity signal in Doppler observations. A much longer time series would be required.

K2-19 presents a rare opportunity to compare the results of the RV method and the TTV method for planetary mass measurement. So far the majority of exoplanets have their mass measured with one or the other of these two methods.  The TTV method seems to be yielding systematically smaller masses than the RV method \citep{WeissMarcy2014}. To illustrate this, we have used different colors in Fig.~\ref{rm} to represent mass measurements based on the TTV and Doppler methods. Many of the planetary systems unveiled by TTV analysis have surprisingly low densities \citep{Lissauer2013,Schmitt2014,Masuda2014,Jontof2014}.  \citet{Steffen2015} attributed the discrepancy to selection effects: for a given planet size, the RV method tends to pick up the more massive planets, while the TTV method is less strongly biased toward massive planets. This is because the proximity to MMR amplifies the TTV signal, thus making smaller planets detectable. \citet{LeeChiang2015} offered a related explanation. Since the TTV method is more sensitive to planets in MMR, and convergent disk migration tends to produce planets in resonant pairs, it is possible that the TTV planets are more likely to have undergone disk migration. Consequently, the TTV planets tend to originate further out in the protoplanetary disk. The outer, thus colder and optically thinner part of the disk is more conducive for the formation of low density planets ("super-puffs"), because the gaseous atmosphere of these planets can readily cool and contract thereby allowing more gas to be accreted. Therefore, TTV planets more likely have lower densities.

As always it is useful to have at least some systems where more than one measurement technique can be applied. The comparison of TTV and RV measurements for a single system has only been achieved in a few cases: Kepler-11 \citep{Lissauer2013,Weiss2015}, WASP-47 \citep{Becker2015,Dai2015} and Kepler-89 \citep{Hirano2012,Masuda2013,Weiss2013}. For the former two systems, the TTV and Doppler methods agree within reported uncertainties. However, for Kepler-89 the TTV mass measurement by \citet{Masuda2013} is lower than the Doppler mass measurements reported by both \citet{Hirano2012} and \citet{Weiss2013}. In the former case the discrepancy is 1$\sigma$ and in the latter case it is nearly 4$\sigma$.

K2-19 offers another opportunity to compare TTV and RV mass determinations. The system has been analyzed with the TTV method by \citet{Narita2015} and \citet{Barros2015}.  Using the analytical TTV model presented by \citet{Deck2015}, \citet{Narita2015} found $M_c$ = $21.4 \pm 1.9~M_{\oplus}$.  This is in agreement with our RV mass: $M_c = 25.6^{+7.1}_{-7.1} ~M_{\oplus}$.  \citet{Barros2015} derived masses for K2-19b and c by analyzing the short-term synodic signal ("chopping") in the TTV of the system with a photodynamical model. \footnote{\citet{Vanderburg2016} and \citet{Sinukoff2015} recently revealed another planet in the system, K2-19d a super-Earth on a 2.5 day orbit. With its presence in the system, the photodynamical model may need to be revised. However, given that planet d is much smaller and far from mean-motion resonance with the other two planets, its effect on other two planets should be quite small.} The results were $M_b$ = $44 \pm 12~M_{\oplus}$ and $M_c$ = $15.9 \pm 7.0~M_{\oplus}$. These results were based on an assumed stellar mass of $0.918^{+0.086}_{-0.070}~M_{\odot}$, whereas we assumed $0.93 \pm 0.05~M_{\odot}$ based on work by \citet{Sinukoff2015}. To facilitate comparison, we rescaled our results under the same assumption for the stellar mass as \citet{Barros2015}, finding the RV masses to be $M_b$ = $28.1^{+5.3}_{-4.9}~M_{\oplus}$ and $M_c$ = $25.3^{+7.0}_{-7.0}~M_{\oplus}$. These differ from the photodynamical masses by 1.2$\sigma$ and 0.9$\sigma$, respectively, for planets b and c.

In addition to the planetary masses, \citet{Barros2015} also reported the mass ratio $\frac{M_c}{M_b} = 0.42 \pm 0.12$. The ratio can be directly constrained based on the observed
synodic chopping signal for mildly eccentric, nearly coplanar systems \citep{Nes2014,Deck2015}. In contrast, in our RV analysis, the mass ratio is relatively poorly constrained ($\frac{M_c}{M_b} = 0.90^{+0.39}_{-0.35}$).

The low mean density ($ 0.334^{+0.081}_{-0.077}$~g~cm$^{-3}$) of K2-19b suggests a large envelope-to-core mass ratio. The low surface gravity of K2-19b coupled with its relatively large size makes it a potentially favorable target for transmission spectroscopy. K2-19c has a size of $4.86^{+0.62}_{-0.44}~R_{\oplus}$ and a mean density of $1.18^{+0.57}_{-0.57}$~g~cm$^{-3}$. The low density suggests that the planet most likely contains a subsantial H/He envelope. Due to the small size of planet d, our current dataset only placed an upper limit on its mass. More observations are needed before we can discuss the nature of planet d.

\citet{Petigura2016} found the masses of K2-24 b and c to be $M_b$ = $21.0 \pm 5.4 ~M_{\oplus}$ and $M_c$ = $27.0 \pm 6.9~M_{\oplus}$. Based on the masses, radii and irradiation levels of the two planets, they were able to estimate the relative envelope-to-core mass ratio by fitting the data with models by \citet{Lopez2014} for the interior compisition and thermal evolution. They found that the two planets have similar core masses: $17.6 \pm 4.3 ~M_{\oplus}$ and $16.1 \pm 4.2 ~M_{\oplus}$, while having different envelope mass fractions of $24 \pm 8 \%$ and $48 \pm 9  \%$. The results of their analysis posed intriguing questions about the formation scenarios of the two planets. How did the planets avoid runaway core accretion with core masses of $\approx$16~$M_{\oplus}$? Why did they end up with such disparate envelope-to-core mass ratios? Our analysis has confirmed and refined the mass measurements of the two planets: $M_b$ = $19.8^{+4.5}_{-4.4} ~M_{\oplus}$ and $M_c$ = $26.0^{+5.8}_{-6.1} ~M_{\oplus}$, and weakened the detection of a constant acceleration term reported by \citet{Petigura2016}. However, it does not alter the basic picture painted by \citet{Petigura2016}. EPIC~205071984b has very similar properties as K2-24b. Again, its low density suggests a substantial gaseous envelope. The upper limits for EPIC~205071984 c and d, $M_c$ < $8.1 ~M_{\oplus}$ and $M_d$ < $35.0 ~M_{\oplus}$ both at a 95\% confidence level, are reasonable for Neptune-sized planets.

The mass of EPIC~204129699b was measured with relatively high precision: $1.857 \pm 0.081~M_{\text{Jup}}$. However, due to the 30~min time averaging of the {\it K2} photometric data and the grazing trajectory of the transiting planet, the uncertainties in the transit parameters are larger than usual \citep{Grziwa2015}. As a result, the planetary radius (and mean density) are relatively poorly constrained. Given that the host star is relatively bright ($V\approx 11$), a priority for future work should be ground-based transit observations with better time sampling, which will help to break the degeneracies.

\acknowledgements We thank Kento Masuda for helpful discussions. We thank Saul Rappaport and Tushar Shrotriya for their contributions to the HARPS observing proposal. We are grateful to Xavier Bonfils, Nicola Astudillo and collaborators, with whom we
exchanged observing time at ESO's 3.6m telescope, facilitating
our observations. We thank Jason Eastman, Emmanuel Jehin and Michaël Gillon for scheduling ground based transit observations with LCOGT and TRAPPIST. Work by FD and JNW was supported by the {\it Transiting
  Exoplanet Survey Satellite} mission. NN acknowledges supports by the NAOJ Fellowship,
Inoue Science Research Award, and Grant-in-Aid for Scientific Research
(A) (JSPS KAKENHI Grant Number 25247026). This work was performed [in part] under contract with the California Institute of Technology (Caltech)/Jet Propulsion Laboratory (JPL) funded by NASA through the Sagan Fellowship Program executed by the NASA Exoplanet Science Institute. AV is supported by the National
Science Foundation Graduate Research Fellowship, Grant No. DGE
1144152. Australian access to the Magellan Telescopes was supported by
the Australian Federal Government through the Department of Industry
and Science.

{\it Facilities:} \facility{Magellan:Clay (Planet Finder
  Spectrograph)}

\bibliography{K2_RV}

\begin{table}
\centering
\caption{System parameters of K2-3}
\begin{tabular}{lrrrr}
\hline
\hline
{\rm Parameter}  & {\rm } & {\rm } &  {\rm }   & {\rm Ref.}  \\
\hline
    Stellar Parameters &$ 	    $&$      $& $$&$   $\\
    $T_{\text{eff}} ~(K)$ &$ 3896 \pm 189     $&$      $& $$& B \\
    $\log~g~(\text{dex})$ &$ 	4.72\pm 0.13  $&$      $& $$& C \\
    $[\text{Fe/H}]~(\text{dex})$ &$ 	-0.32 \pm 0.13  $&$      $& $$& B \\
    $M_{\star} ~(~M_{\odot})$ &$ 0.601 \pm 0.089     $&$      $& $$& B \\
    $R_{\star} ~(R_{\odot})$ &$ 0.561 \pm 0.068     $&$      $& $$& B \\
    $\text{Apparent V mag}$ &$ 12.17 \pm 0.01     $&$      $& $$& A \\
    \\
    Planetary Parameters &$ b	    $&$   c   $& $d^\star$&$   $\\   
    Transit Model\\ 
    $P ~(\text{days})$ &$ 10.05403^{+0.0026}_{-0.0025}   $&$ 24.6454^{+0.0013}_{-0.0013}    $& $ 44.5631^{+0.0063}_{-0.0055} $& B\\
    $t_0 ~(\text{BJD-2454900})$ &$ 1913.4189^{+0.0011}_{-0.0011}   $&$ 1912.2786^{+0.0026}_{-0.0027}     $& $ 1926.2232^{+0.0037}_{-0.0043} $& B\\
    $R_{p} ~(R_{\oplus})$ &$ 2.14^{+0.27}_{-0.26}      $&$ 1.72^{+0.23}_{-0.22}      $& $1.52^{+0.21}_{-0.20} $& B\\
    \\
    Circular RV model \\
    $K$ ~(m~s$^{-1})$ &$ 3.36^{+0.75}_{-0.74}       $&$ <1.3~(95\%~conf.)    $& $1.91^{+0.74}_{-0.74}$& A \\
    $M_{p} ~(~M_{\oplus})$ &$ 8.1^{+2.0}_{-1.9}     $&$ < 4.2~(95\%~conf.)    $& $7.5^{+3.0}_{-3.0}$& A \\
    $\rho$ ~(g~cm$^{-3})$ &$  4.5^{+2.0}_{-2.0}    $&$   < 4.5~(95\%~conf.)   $& $ 11.7^{+6.7}_{-6.7}  $& A \\
     $\sigma_{\text{jit},~\text{PFS}}$ ~(m~s$^{-1})$ &$ 5.7^{+1.0}_{-0.8}    $&$     $& $$& A \\
     $\sigma_{\text{jit},~\text{HARPS}}$ ~(m~s$^{-1})$ &$2.97^{+0.58}_{-0.53}     $&$     $& $$& A \\

 \\    
         Eccentric RV model \\
    $K$ ~(m~s$^{-1})$ &$ 3.26^{+0.81}_{-0.79}       $&$  <3.9~(95\%~conf.)     $& $2.9^{+1.5}_{-1.4}$& A \\
    $M_{p} ~(~M_{\oplus})$ &$ 7.7^{+2.0}_{-2.0}     $&$ <12.6~(95\%~conf.)     $& $11.3^{+5.9}_{-5.8}$& A \\
    $\rho$ ~(g~cm$^{-3})$ &$  4.3^{+2.0}_{-2.0}    $&$ <13.5~(95\%~conf.)      $& $ 18^{+12}_{-12}  $& A \\
     $e$ &$  0.21^{+0.15}_{-0.12}    $&$ Unconstrained     $& $ 0.65^{+0.17}_{-0.43}  $& A \\
     $\sigma_{\text{jit},~\text{PFS}}$ ~(m~s$^{-1})$ &$ 5.4^{+1.1}_{-0.9}    $&$     $& $$& A \\
     $\sigma_{\text{jit},~\text{HARPS}}$ ~(m~s$^{-1})$ &$2.74^{+0.65}_{-0.63}     $&$     $& $$& A \\
     
   \\
   
       Circular RV model including stellar activity\\
    $K$ ~(m~s$^{-1})$ &$ 2.87^{+0.72}_{-0.67}       $&$ 1.60^{+0.80}_{-0.91}    $& $5.0^{+1.7}_{-1.6}$& A \\
    $M_{p} ~(~M_{\oplus})$ &$ 6.9^{+1.9}_{-1.7}     $&$ 5.2^{+2.6}_{-3.0}     $& $19.8^{+6.8}_{-6.6}$& A \\
    $\rho$ ~(g~cm$^{-3})$ &$  3.8^{+1.8}_{-1.7}    $&$ 5.6^{+3.6}_{-3.9}     $& $ 31^{+17}_{-16}  $& A \\
     $\sigma_{\text{jit},~\text{PFS}}$ ~(m~s$^{-1})$ &$ 5.0^{+1.7}_{-1.6}    $&$     $& $$& A \\
     $\sigma_{\text{jit},~\text{HARPS}}$ ~(m~s$^{-1})$ &$2.16^{+0.62}_{-0.64}     $&$     $& $$& A \\
     $\text{Period of stellar activity}$ ~(days) &$39.7^{+0.9}_{-1.0}     $&$     $& $$& A \\
     $\text{Amplitude of stellar activity}$ ~(m~s$^{-1})$ &$3.9 ^{+1.4}_{-1.5}    $&$     $& $$& A \\
     $\text{Phase of stellar activity}$ ~(rad) &$ 2.1 ^{+2.8}_{-1.6}    $&$     $& $$& A \\
\hline
\end{tabular}
\tablecomments{A: This Work; B: \citet{Crossfield2015}; C: \citet{Sinukoff2015}. $\star$: The radial velocity result for planet d is highly degenerate with stellar activity signal. See text for details.}
\label{201367065para}
\end{table}

\begin{table}
\centering
\caption{System parameters of K2-19}
\begin{tabular}{lrrrr}
\hline
\hline
{\rm Parameter}  & {\rm } & {\rm } &  {\rm }   & {\rm Ref.}  \\
\hline
    Stellar Parameters &$ 	    $&$      $& $$&$   $\\
    $T_{\text{eff}} ~(K)$ &$ 5430\pm60     $&$      $& $$& B \\
    $\log~g~(\text{dex})$ &$ 	4.63\pm 0.07  $&$      $& $$& B \\
    $[\text{Fe/H}]~(\text{dex})$ &$ 	0.10\pm 0.04  $&$      $& $$& B \\
    $v~\text{sin}~i$ ~(km~s$^{-1})$ &$ < 2     $&$      $& $$& B \\
    $M_{\star} ~(~M_{\odot})$ &$ 0.93 \pm 0.05     $&$      $& $$& B \\
    $R_{\star} ~(R_{\odot})$ &$ 0.86 \pm 0.04     $&$      $& $$& B \\
    $\text{Apparent V mag}$ &$ 13.00 \pm 0.01     $&$      $& $$& B \\
    \\
    Planetary Parameters &$ b	    $&$ c     $& $d$&$   $\\    
    Transit Model\\
    $P ~(\text{days})$ &$ 7.91940 \pm 0.00005    $&$ 11.90715 \pm 0.00150   $& $ 2.50856 \pm 0.00041 $& B\\
    $t_0 ~(\text{BJD-2454900})$ &$ 1913.3837 \pm 0.0003     $&$ 1917.2755\pm 0.0051    $& $ 1908.9207 \pm 0.0086$& B\\
        $R_{p} ~(R_{\oplus})$ &$ 7.74\pm 0.39     $&$ 4.86^{+0.62}_{-0.44}    $& $ 1.14 \pm 0.13$& B\\
    \\
    Circular RV model (with 12-day averaging)\\

    $K$ ~(m~s$^{-1})$ &$ 9.6^{+1.8}_{-1.6}      $&$ 7.5^{+2.1}_{-2.1}    $& $ <6.9$~(95\%~conf.)& A \\
    $M_{p} ~(~M_{\oplus})$ &$ 28.5^{+5.4}_{-5.0}     $&$ 25.6^{+7.1}_{-7.1}     $& $ < 14.0 $~(95\%~conf.)& A \\
    $\rho$ ~(g~cm$^{-3})$ &$  0.334^{+0.081}_{-0.077}    $&$   1.18^{+0.57}_{-0.57}    $& $ < 51.4$~(95\%~conf.)& A \\
     $\sigma_{\text{jit},~\text{PFS}}$ ~(m~s$^{-1})$ &$ 10.0^{+1.4}_{-1.2}    $&$     $& $$& A \\
     $\sigma_{\text{jit},~\text{HARPS}}$ ~(m~s$^{-1})$ &$2.2^{+2.8}_{-1.7}     $&$     $& $$& A \\
     
         \\
         Circular RV model (without 12-day averaging)\\

    $K$ ~(m~s$^{-1})$ &$ 9.0^{+1.7}_{-1.6}      $&$ 6.3^{+1.9}_{-2.0}    $& $ <7.7$~(95\%~conf.)& A \\
    $M_{p} ~(~M_{\oplus})$ &$ 26.2^{+5.0}_{-4.8}     $&$ 20.9^{+6.5}_{-6.6}     $& $ < 15.6 $~(95\%~conf.)& A \\
    $\rho$ ~(g~cm$^{-3})$ &$  0.307^{+0.075}_{-0.072}    $&$   0.96^{+0.52}_{-0.53}    $& $ < 57.3$~(95\%~conf.)& A \\
     $\sigma_{\text{jit},~\text{PFS}}$ ~(m~s$^{-1})$ &$ 11.8^{+1.5}_{-1.2}    $&$     $& $$& A \\
     $\sigma_{\text{jit},~\text{HARPS}}$ ~(m~s$^{-1})$ &$1.9^{+2.3}_{-1.3}     $&$     $& $$& A \\
     \\
    Eccentric RV model\\ 

    $K$ ~(m~s$^{-1})$ &$ 11.42^{+2.3}_{-2.4}      $&$ 7.8^{+2.9}_{-3.2}    $& $ <9.7$~(95\%~conf.)& A \\
    $M_{p} ~(~M_{\oplus})$ &$ 31.8^{+6.7}_{-7.0}     $&$ 26.5^{+9.8}_{-10.8}     $& $<9.6 $~(95\%~conf.)& A \\
    $\rho$ ~(g~cm$^{-3})$ &$  0.374^{+0.095}_{-0.089}    $&$   1.26^{+0.67}_{-0.70}    $& $  <35.2$ ~(95\%~conf.) & A \\
    $e$ &$  < 0.66 $~(95\%~conf.)   &$  < 0.82$ ~(95\%~conf.)   & $  Unconstrained  $& A \\
     $\sigma_{\text{jit},~\text{PFS}}$ ~(m~s$^{-1})$ &$ 9.7^{+2.4}_{-1.5}    $&$     $& $$& A \\
     $\sigma_{\text{jit},~\text{HARPS}}$ ~(m~s$^{-1})$ &$2.2^{+3.1}_{-1.6}     $&$     $& $$& A \\
\hline
\end{tabular}
\tablecomments{A: This work. B: \citet{Sinukoff2015}}
\label{201505350para}
\end{table}

\begin{table}
\centering
\caption{System parameters of K2-24}
\begin{tabular}{lrrr}
\hline
\hline
{\rm Parameter}  & {\rm } & {\rm }   & {\rm Ref.}  \\
\hline
    Stellar Parameters &$ 	    $&$      $&$   $\\
    $T_{\text{eff}} ~(K)$ &$ 5743 \pm 60     $&$      $& B \\
    $\log~g ~(\text{dex})$ &$ 	4.29 \pm 0.07  $&$      $& B \\
    $[\text{Fe/H}] ~(\text{dex})$ &$ 	0.42 \pm 0.04  $&$      $& B \\
    $v~\text{sin}~i$ ~(km~s$^{-1})$ &$ <2     $&$      $& B \\
    $M_{\star} ~(~M_{\odot})$ &$ 1.12 \pm 0.05     $&$      $& B \\
    $R_{\star} ~(R_{\odot})$ &$ 1.21 \pm 0.11     $&$      $& B \\
    $\text{Apparent V mag}$ &$ 11.07 \pm 0.13     $&$      $& B \\
    \\
    Planetary Parameters &$ b	    $&$  c    $&$   $\\   
    Transit Model \\ 
    $P ~(\text{days})$ &$ 20.8851 \pm 0.0003   $&$ 42.3633 \pm 0.0006    $&  B\\
    $t_0 ~(\text{BJD-2454900})$ &$ 2005.7948 \pm 0.0007   $&$ 2015.6251 \pm 0.0004    $& B\\
    $R_{p} ~(R_{\oplus})$ &$ 5.68 \pm 0.56      $&$ 7.82 \pm 0.72      $& B\\
    \\
    Circular RV model\\
    $K$ ~(m~s$^{-1})$ &$ 4.25^{+0.95}_{-0.94}       $&$ 4.4^{+1.0}_{-1.0}    $& A \\
    $M_{p} ~(~M_{\oplus})$ &$ 19.8^{+4.5}_{-4.4}     $&$ 26.0^{+5.8}_{-6.1}     $ &A \\ 
    $\rho$ ~(g~cm$^{-3})$ &$  0.59 \pm 0.22    $&$ 0.30^{+0.11}_{-0.11}     $& A \\
     $\dot \gamma$ ~(m~s$^{-1}$~yr$^{-1})$ &$  -12^{+10}_{-10}    $&$    $& A \\  
     $\sigma_{\text{jit},~\text{PFS}}$ ~(m~s$^{-1})$ &$ 3.7^{+1.0}_{-0.8}    $& $$& A \\
     $\sigma_{\text{jit},~\text{HARPS}}$ ~(m~s$^{-1})$ &$3.9^{+2.1}_{-1.8}     $& $$& A \\
     $\sigma_{\text{jit},~\text{HIRES}}$ ~(m~s$^{-1})$ &$3.57^{+0.75}_{-0.59}     $& $$& A \\
     \\
    Eccentric RV model\\
    $K$ ~(m~s$^{-1})$ &$ 5.0^{+1.0}_{-1.0}      $&$ 5.4^{+1.3}_{-1.1}    $& A \\
    $M_{p} ~(~M_{\oplus})$ &$ 23.3^{+4.8}_{-4.7}     $&$ 30.7^{+7.1}_{-6.3}     $ &A \\ 
    $\rho$ ~(g~cm$^{-3})$ &$  0.69 \pm 0.25    $&$ 0.35^{+0.13}_{-0.12}   $& A \\
    $e$ &$  0.24^{+0.10}_{-0.11}    $&$ <0.58$~(95\%~conf.)     & A \\
     $\dot \gamma$ ~(m~s$^{-1}$~yr$^{-1})$ &$  -14.9^{+8.3}_{-8.1}    $&$    $& A \\  
     $\sigma_{\text{jit},~\text{PFS}}$ ~(m~s$^{-1})$ &$ 3.43^{+0.96}_{-0.71}    $& $$& A \\
     $\sigma_{\text{jit},~\text{HARPS}}$ ~(m~s$^{-1})$ &$3.6^{+2.0}_{-1.6}     $& $$& A \\
     $\sigma_{\text{jit},~\text{HIRES}}$ ~(m~s$^{-1})$ &$3.13^{+0.67}_{-0.57}     $& $$& A \\
\hline
\end{tabular}
\tablecomments{A: This Work; B: \citet{Petigura2016}.}
\label{203771098para}
\end{table}

\begin{table}
\centering
\caption{System parameters of EPIC~204129699}
\begin{tabular}{lrr}
\hline
\hline
{\rm Parameter}  & {\rm }   & {\rm Ref.}  \\
\hline
    Stellar Parameters &$ 	    $&$   $\\
    $T_{\text{eff}} ~(K)$ &$ 5412 \pm 34     $& A \\
    $\log~g ~(\text{dex})$ &$ 	4.44\pm  0.05  $& A \\
    $[\text{Fe/H}] ~(\text{dex})$ &$ 	0.20\pm0.03  $& A \\
    $v~\text{sin}~i$ ~(km~s$^{-1})$ &$ 2.6_{-1.0}^{+1.0}     $& A \\
    $M_{\star} ~(~M_{\odot})$ &$  1.000 \pm 0.064    $& A \\
    $R_{\star} ~(R_{\odot})$ &$ 0.986 \pm 0.070     $& A \\
    $\text{Apparent V mag}$ &$ 10.775 \pm   0.023  $& A \\
    \\
    Planetary Parameters & $b$&$   $\\    
    Transit Model\\
    $P ~(\text{days})$ &$ 1.257850 \pm 0.000002  $& B\\
    $t_0 ~(\text{BJD-2454900})$ &$ 2291.70889 \pm 0.00024   $& B\\
    $R_{p} ~(R_{\oplus})$ & $Unconstrained$& B\\
    \\
    Circular RV model\\
    $K$ ~(m~s$^{-1})$ &$ 349.5^{+3.1}_{-3.7}       $& A \\
    $M_{p} ~(~M_{\oplus})$ &$ 590.1^{+25.7}_{-25.9}     $& A \\
    $\rho$ ~(g~cm$^{-3})$ &$ Unconstrained  $& A \\
     $\sigma_{\text{jit},~\text{PFS}}$ ~(m~s$^{-1})$ &$ 14.9^{+8.6}_{-4.2}    $& A \\
     $\sigma_{\text{jit},~\text{HARPS}}$ ~(m~s$^{-1})$ &$9^{+24}_{-6}     $& A \\
     $\sigma_{\text{jit},~\text{TRES}}$ ~(m~s$^{-1})$ &$28^{+12}_{-8}     $& A \\
     $\sigma_{\text{jit},~\text{FIES}}$ ~(m~s$^{-1})$ &$3.3^{+5.1}_{-2.3}     $& A \\
     \\
      Eccentric RV model\\
    $K$ ~(m~s$^{-1})$ &$ 348.5^{+3.4}_{-3.6}       $& A \\
    $M_{p} ~(~M_{\oplus})$ &$ 564.2^{+8.3}_{-9.5}     $& A \\
    $\rho$ ~(g~cm$^{-3})$ &$ Unconstrained $& A \\
     $e$ &$ <0.027~(95\%~conf.)      $& A \\
     $\sigma_{\text{jit},~\text{PFS}}$ ~(m~s$^{-1})$ &$ 16^{+11}_{-5}    $& A \\
     $\sigma_{\text{jit},~\text{HARPS}}$ ~(m~s$^{-1})$ &$7^{+20}_{-5}     $& A \\
     $\sigma_{\text{jit},~\text{TRES}}$ ~(m~s$^{-1})$ &$29^{+12}_{-8}     $& A \\
     $\sigma_{\text{jit},~\text{FIES}}$ ~(m~s$^{-1})$ &$3.3^{+5.3}_{-2.4}     $& A \\  
     
\hline
\end{tabular}
\tablecomments{A: This Work; B: \citet{Grziwa2015}.}
\label{204129699para}
\end{table}

\begin{table}
\centering
\caption{System parameters of EPIC~205071984}
\begin{tabular}{lrrrr}
\hline
\hline
{\rm Parameter}  & {\rm } & {\rm } &  {\rm }   & {\rm Ref.}  \\
\hline
    Stellar Parameters &$ 	    $&$      $& $$&$   $\\
    $T_{\text{eff}} ~(K)$ &$ 5315\pm60     $&$      $& $$& B \\
    $\log~g ~(\text{dex})$ &$ 	4.43\pm 0.07  $&$      $& $$& B \\
    $[\text{Fe/H}] ~(\text{dex})$ &$ 	0.00\pm0.04  $&$      $& $$& B \\
    $v~\text{sin}~i$ ~(km~s$^{-1})$ &$ <2     $&$      $& $$& B \\
    $M_{\star} ~(~M_{\odot})$ &$ 0.87 \pm 0.04     $&$      $& $$& B \\
    $R_{\star} ~(R_{\odot})$ &$ 0.87 \pm 0.05     $&$      $& $$& B \\
    $\text{Apparent V mag}$ &$ 12.332 \pm 0.017     $&$      $& $$& B \\
    \\
    Planetary Parameters &$ 	b    $&$    c  $& $d$&$   $\\    
    Transit Model\\
    $P ~(\text{days})$ &$ 8.99218 \pm 0.00020    $&$ 20.65614 \pm 0.00598   $& $ 31.71922 \pm 0.00236 $& B\\
    $t_0 ~(\text{BJD-2454900})$ &$ 2000.9258 \pm 0.0009     $&$ 1999.4306\pm 0.010    $& $ 903.7846 \pm 0.0031$& B\\
    $R_{p} ~(R_{\oplus})$ &$ 5.38\pm 0.35     $&$ 3.48^{+0.97}_{-0.42}    $& $3.75 \pm 0.40$& B\\
    \\
     Circular  RV model\\
    $K$ ~(m~s$^{-1})$ &$ 7.1^{+2.0}_{-2.0}      $&$ <2.1$~(95\%~conf.)   & $<7.8$~(95\%~conf.) & A \\
    $~M_{p} ~(~M_{\oplus})$ &$ 21.1^{+5.9}_{-5.9}     $&$ <8.1$~(95\%~conf.)     & $<35.0$~(95\%~conf.)& A \\

    $\rho$ ~(g~cm$^{-3})$ &$ 0.74^{+ 0.25}_{-0.25}    $&$  <1.1 $~(95\%~conf.)  & $ <3.6$~(95\%~conf.)& A \\
        $\dot \gamma$ ~(m~s$^{-1}$~yr$^{-1})$ &$  34.0^{+9.9}_{-9.7}    $&$    $&$$& A \\ 
     $\sigma_{\text{jit},~\text{PFS}}$ ~(m~s$^{-1})$ &$ 5.4^{+3.5}_{-2.1}    $&$     $& $$& A \\
     $\sigma_{\text{jit},~\text{HARPS}}$ ~(m~s$^{-1})$ &$3.37^{+0.69}_{-0.58}     $&$     $& $$& A \\
     \\
   Eccentric RV model  &$ 	    $&$      $& $$&$   $\\     
       $K$ ~(m~s$^{-1})$ &$ 8.6^{+2.6}_{-2.4}      $&$ <4.3$~(95\%~conf.)   & $ < 8.9$~(95\%~conf.)& A \\
    $M_{p} ~(~M_{\oplus})$ &$ 25.0^{+7.4}_{-7.0}     $&$ <  16.5 $~(95\%~conf.)   & $ <40.3$~(95\%~conf.)& A \\

    $\rho$ ~(g~cm$^{-3})$ &$ 0.88^{+0.31}_{-0.30}    $&$  < 2.1$~(95\%~conf.) & $  <4.2$~(95\%~conf.)& A \\
    $e$ &$   <  0.43$~(95\%~conf.)    &$ Unconstrained   $& $ Unconstrained $& A \\
        $\dot \gamma$ ~(m~s$^{-1}$ yr$^{-1})$ &$  34^{+13}_{-13}    $&$    $&$$& A \\ 
     $\sigma_{\text{jit},~\text{PFS}}$ ~(m~s$^{-1})$ &$ 7.2^{+4.3}_{-2.8}    $&$     $& $$& A \\
     $\sigma_{\text{jit},~\text{HARPS}}$ ~(m~s$^{-1})$ &$3.26^{+0.75}_{-0.62}     $&$     $& $$& A \\
\hline
\end{tabular}
\tablecomments{A: This work; B: \citet{Sinukoff2015}}
\label{205071984para}
\end{table}

\begin{table}
\centering
\caption{Relative radial velocity of K2-3}
\begin{tabular}{lrrr}
\hline
\hline
{\rm BJD}  & {\rm RV (m~s$^{-1}$)} & {\rm Unc.\ (m~s$^{-1}$)}   & {\rm Source}  \\
\hline
 2457050.75983 &$     -8.34 $&$      2.51 $&$  0 $\\
 2457051.80129 &$      0.00 $&$      2.35 $&$  0 $\\
 2457052.84973 &$     -0.10 $&$      2.61 $&$  0 $\\
 2457054.80341 &$      2.80 $&$      4.38 $&$  0 $\\
 2457055.83610 &$     -3.97 $&$      2.81 $&$  0 $\\
 2457061.76506 &$     13.48 $&$      3.02 $&$  0 $\\
 2457062.82087 &$     -2.22 $&$      2.33 $&$  0 $\\
 2457064.74697 &$     -4.85 $&$      2.63 $&$  0 $\\
 2457065.75519 &$     -5.97 $&$      2.57 $&$  0 $\\
 2457066.72884 &$     -9.21 $&$      2.99 $&$  0 $\\
 2457067.74709 &$    -10.36 $&$      2.74 $&$  0 $\\
 2457068.74840 &$     -8.44 $&$      2.14 $&$  0 $\\
 2457069.75326 &$      1.86 $&$      2.51 $&$  0 $\\
 2457118.61730 &$      0.27 $&$      2.34 $&$  0 $\\
 2457118.62796 &$      0.72 $&$      2.56 $&$  0 $\\
 2457118.63900 &$      3.64 $&$      2.64 $&$  0 $\\
 2457119.62004 &$      4.87 $&$      1.95 $&$  0 $\\
 2457119.63088 &$     -2.53 $&$      2.24 $&$  0 $\\
 2457119.64192 &$      1.86 $&$      2.23 $&$  0 $\\
 2457120.60464 &$     17.54 $&$      1.91 $&$  0 $\\
 2457120.61554 &$      6.90 $&$      2.02 $&$  0 $\\
 2457120.62655 &$      1.24 $&$      1.88 $&$  0 $\\
 2457121.58823 &$      0.75 $&$      2.65 $&$  0 $\\
 2457121.59876 &$     -1.31 $&$      2.55 $&$  0 $\\
 2457121.61042 &$     -3.45 $&$      2.46 $&$  0 $\\
 2457122.59521 &$     -2.62 $&$      2.26 $&$  0 $\\
 2457122.60651 &$      4.05 $&$      2.21 $&$  0 $\\
 2457122.61786 &$     -3.06 $&$      2.59 $&$  0 $\\
 2457123.60398 &$      4.08 $&$      2.56 $&$  0 $\\
 2457123.61534 &$     -4.13 $&$      2.22 $&$  0 $\\
 2457123.62646 &$      0.31 $&$      2.19 $&$  0 $\\
\hline
\end{tabular}
\tablecomments{0: PFS}
\label{rv_table}
\end{table}
 
\begin{table}
\caption{Relative radial velocity of K2-19}
\begin{tabular}{lrrrr}
\hline
\hline
{\rm BJD}  & {\rm RV (m~s$^{-1}$)} & {\rm Unc.\ (m~s$^{-1}$)}   &{\rm BIS.\ (m~s$^{-1}$)}  & {\rm Source} \\
\hline
 2457050.85856 &$     -8.88 $&$      5.19 $& &$  0 $\\
 2457051.81505 &$    -15.46 $&$      3.42 $& &$  0 $\\
 2457052.86521 &$    -29.96 $&$      4.76 $& &$  0 $\\
 2457054.81634 &$      0.96 $&$      6.52 $& &$  0 $\\
 2457055.85049 &$      3.82 $&$      4.55 $& &$  0 $\\
 2457061.70681 &$     20.63 $&$      6.77 $& &$  0 $\\
 2457061.79848 &$     17.55 $&$      4.50 $& &$  0 $\\
 2457061.81294 &$     30.88 $&$      5.05 $& &$  0 $\\
 2457062.77601 &$      6.73 $&$      4.57 $& &$  0 $\\
 2457062.79007 &$     13.98 $&$      4.42 $& &$  0 $\\
 2457062.80443 &$     10.52 $&$      5.10 $& &$  0 $\\
 2457064.76307 &$      8.89 $&$      4.15 $& &$  0 $\\
 2457064.77760 &$     17.62 $&$      4.27 $& &$  0 $\\
 2457064.79190 &$     30.78 $&$      4.92 $& &$  0 $\\
 2457065.76985 &$     33.43 $&$      4.99 $& &$  0 $\\
 2457065.78424 &$     23.95 $&$      4.94 $& &$  0 $\\
 2457065.79889 &$     30.89 $&$      4.83 $& &$  0 $\\
 2457066.75799 &$     26.22 $&$      5.42 $& &$  0 $\\
 2457066.77275 &$     30.65 $&$      6.15 $& &$  0 $\\
 2457066.78723 &$      4.93 $&$      7.64 $& &$  0 $\\
 2457067.76192 &$    -10.04 $&$      5.16 $& &$  0 $\\
 2457067.77624 &$    -22.25 $&$      5.54 $& &$  0 $\\
 2457067.79082 &$     -0.95 $&$      5.19 $& &$  0 $\\
 2457068.76326 &$     14.88 $&$      5.04 $& &$  0 $\\
 2457068.77757 &$      2.66 $&$      5.19 $& &$  0 $\\
 2457068.79244 &$     -6.31 $&$      5.14 $& &$  0 $\\
 2457069.76793 &$    -23.29 $&$      5.11 $& &$  0 $\\
 2457069.78244 &$    -11.48 $&$      5.13 $& &$  0 $\\
 2457069.79654 &$    -12.98 $&$      5.01 $& &$  0 $\\
 2457118.65185 &$    -13.28 $&$      4.09 $& &$  0 $\\
 2457118.66628 &$     -1.10 $&$      4.78 $& &$  0 $\\
 2457118.68049 &$      0.00 $&$      4.24 $& &$  0 $\\
 2457119.65552 &$    -10.47 $&$      3.85 $& &$  0 $\\
 2457119.67002 &$     -7.98 $&$      3.89 $& &$  0 $\\
 2457119.68417 &$     10.92 $&$      4.36 $& &$  0 $\\
 2457120.63937 &$      6.03 $&$      3.72 $& &$  0 $\\
 2457120.65355 &$      7.85 $&$      5.47 $& &$  0 $\\
 2457121.62451 &$      0.24 $&$      4.29 $& &$  0 $\\
 2457121.63942 &$      8.99 $&$      3.82 $& &$  0 $\\
 2457121.65395 &$     19.60 $&$      3.88 $& &$  0 $\\
 2457122.62934 &$      6.10 $&$      5.34 $& &$  0 $\\
 2457122.64080 &$    -16.70 $&$      4.84 $& &$  0 $\\
 2457122.65206 &$     -6.75 $&$      5.11 $& &$  0 $\\
 2457123.63854 &$    -16.07 $&$      5.56 $& &$  0 $\\
 2457123.66345 &$      4.63 $&$      6.12 $& &$  0 $\\
 2457123.67839 &$     -5.18 $&$      5.88 $& &$  0 $\\
 2457124.67522 &$    -21.64 $&$      8.47 $& &$  0 $\\
\hline
\end{tabular}
\begin{tabular}{lrrrr}
\hline
\hline
{\rm BJD}  & {\rm RV (m~s$^{-1}$)} & {\rm Unc.\ (m~s$^{-1}$)}   &{\rm BIS.\ (m~s$^{-1}$)}  & {\rm Source} \\
\hline
 2457185.51808 &$      3.74 $&$      3.19 $&$$&$  1 $\\
 2457186.50677 &$      0.01 $&$      4.83 $&$41.05$&$  1 $\\
 2457187.53841 &$     -4.75 $&$      3.55 $&$43.01$&$  1 $\\
 2457188.54280 &$     -8.46 $&$      3.50 $&$29.47$&$  1 $\\
 2457189.57695 &$     -1.20 $&$      3.56 $&$16.00$&$  1 $\\
 2457190.57602 &$     -1.17 $&$      3.11 $&$37.20$&$  1 $\\
 2457191.59452 &$     11.82 $&$      3.99 $&$28.18$&$  1 $\\
 2457203.50361 &$    -11.02 $&$      4.73 $&$29.63$&$  1 $\\
 2457198.47072 &$    -10.79 $&$      5.13 $& &$  0 $\\
 2457198.52715 &$     -6.69 $&$      4.53 $& &$  0 $\\
 2457199.51333 &$    -15.31 $&$      4.49 $& &$  0 $\\
 2457199.52804 &$     -1.28 $&$      5.21 $& &$  0 $\\
 2457200.52892 &$     -9.33 $&$      5.14 $& &$  0 $\\
 2457200.54451 &$     10.03 $&$      7.79 $& &$  0 $\\
 2457203.52075 &$    -29.66 $&$      5.43 $& &$  0 $\\
 2457203.53537 &$    -47.19 $&$      6.68 $& &$  0 $\\
 2457204.49376 &$     -5.84 $&$      4.48 $& &$  0 $\\
 2457204.50816 &$    -20.38 $&$      4.80 $& &$  0 $\\
 2457205.50352 &$    -20.85 $&$      4.69 $& &$  0 $\\
 2457205.51810 &$     -5.42 $&$      4.48 $& &$  0 $\\
 2457206.49605 &$     11.04 $&$      4.13 $& &$  0 $\\
 2457206.50796 &$     10.19 $&$      4.49 $& &$  0 $\\
\hline
\end{tabular}
\tablecomments{0: PFS; 1: HARPS}
\label{rv_table}
\end{table}
 
\begin{table}
\centering
\caption{Relative radial velocity of K2-24}
\begin{tabular}{lrrrr}
\hline
\hline
{\rm BJD}  & {\rm RV (m~s$^{-1}$)} & {\rm Unc.\ (m~s$^{-1}$)} &{\rm BIS.\ (m~s$^{-1}$)}   & {\rm Source}  \\
\hline
 2457198.61435 &$      9.03 $&$      1.79 $&&$  0 $\\
 2457198.69058 &$      9.35 $&$      1.91 $&&$  0 $\\
 2457199.60735 &$      3.53 $&$      1.77 $&&$  0 $\\
 2457199.76091 &$      5.34 $&$      1.79 $&&$  0 $\\
 2457200.70251 &$     -3.16 $&$      1.43 $&&$  0 $\\
 2457200.71696 &$      4.18 $&$      1.56 $&&$  0 $\\
 2457202.64632 &$      0.00 $&$      1.75 $&&$  0 $\\
 2457202.66105 &$     -5.46 $&$      1.91 $&&$  0 $\\
 2457203.55168 &$      8.16 $&$      1.43 $&&$  0 $\\
 2457203.56610 &$      2.72 $&$      1.52 $&&$  0 $\\
 2457204.67286 &$     -3.11 $&$      1.54 $&&$  0 $\\
 2457204.68711 &$     -0.53 $&$      1.53 $&&$  0 $\\
 2457205.60288 &$      0.87 $&$      1.54 $&&$  0 $\\
 2457205.61494 &$     -0.73 $&$      1.57 $&&$  0 $\\
 2457206.63194 &$     -1.82 $&$      1.86 $&&$  0 $\\
 2457206.64440 &$     -0.84 $&$      1.69 $&&$  0 $\\
 2457190.62372 &$      1.86 $&$      1.33 $&23.05&$  1 $\\
 2457190.81012 &$      3.51 $&$      1.68 $&35.98&$  1 $\\
 2457191.63661 &$     -2.34 $&$      2.55 $&19.09&$  1 $\\
 2457191.64804 &$      4.43 $&$      2.41 $&21.81&$  1 $\\
 2457249.50031 &$     -4.52 $&$      3.79 $&37.23&$  1 $\\
 2457253.46724 &$      4.10 $&$      1.52 $&37.28&$  1 $\\
 2457266.49748 &$     -7.77 $&$      2.48 $&32.82&$  1 $\\
 2457267.54626 &$     -3.18 $&$      1.56 $&22.12&$  1 $\\
 2457270.48326 &$     -1.91 $&$      1.47 $&41.09&$  1 $\\
 2457277.47435 &$      5.83 $&$      1.86 $&26.13&$  1 $\\
\hline
\end{tabular}
\tablecomments{0: PFS; 1: HARPS}
\label{rv_table}
\end{table}
 
\begin{table}
\centering
\caption{Relative radial velocity of EPIC 204129699}
\begin{tabular}{lrrr}
\hline
\hline
{\rm BJD}  & {\rm RV (m~s$^{-1}$)} & {\rm Unc.\ (m~s$^{-1}$)}   & {\rm Source}  \\
\hline
 2457199.59189 &$   -284.80 $&$      1.64 $&$  0 $\\
 2457201.60623 &$    289.83 $&$      2.22 $&$  0 $\\
 2457202.63203 &$    368.56 $&$      1.33 $&$  0 $\\
 2457203.63081 &$      0.00 $&$      1.34 $&$  0 $\\
 2457204.65786 &$   -277.36 $&$      1.21 $&$  0 $\\
 2457205.59027 &$    -24.70 $&$      1.28 $&$  0 $\\
 2457206.61979 &$    347.99 $&$      1.46 $&$  0 $\\
 2457122.88109 &$   -487.18 $&$     13.52 $&$  2 $\\
 2457144.86865 &$    199.73 $&$     16.65 $&$  2 $\\
 2457145.82832 &$   -132.03 $&$     14.09 $&$  2 $\\
 2457146.83816 &$   -449.60 $&$     14.24 $&$  2 $\\
 2457148.89470 &$    -63.91 $&$     14.03 $&$  2 $\\
 2457149.88110 &$    224.15 $&$     24.69 $&$  2 $\\
 2457150.88949 &$     -3.44 $&$     16.81 $&$  2 $\\
 2457151.88311 &$   -410.65 $&$     14.66 $&$  2 $\\
 2457152.83368 &$   -340.50 $&$     14.12 $&$  2 $\\
 2457153.85085 &$      0.00 $&$     13.52 $&$  2 $\\
\hline
\end{tabular}
\tablecomments{0: PFS; 2: TRES}
\label{rv_table}
\end{table}
 
\begin{table}
\centering
\caption{Relative radial velocity of EPIC 205071984}
\begin{tabular}{lrrrr}
\hline
\hline
{\rm BJD}  & {\rm RV (m~s$^{-1}$)} & {\rm Unc.\ (m~s$^{-1}$)}&{\rm BIS.\ (m~s$^{-1}$)}   & {\rm Source}  \\
\hline
 2457198.67460 &$    -13.95 $&$      2.31 $&&$  0 $\\
 2457199.74554 &$    -24.28 $&$      2.56 $&&$  0 $\\
 2457200.68722 &$    -12.24 $&$      2.16 $&&$  0 $\\
 2457204.72413 &$      0.00 $&$      2.40 $&&$  0 $\\
 2457205.64081 &$      2.07 $&$      2.09 $&&$  0 $\\
 2457206.66905 &$      2.37 $&$      2.67 $&&$  0 $\\
 2457185.60690 &$     10.69 $&$      2.65 $&14.16&$  1 $\\
 2457185.67718 &$     10.92 $&$      2.34 $&8.36&$  1 $\\
 2457185.81974 &$      9.38 $&$      2.73 $&1.37&$  1 $\\
 2457185.84975 &$      3.16 $&$      4.14 $&1.26&$  1 $\\
 2457186.59974 &$      1.56 $&$      3.49 $&7.83&$  1 $\\
 2457186.66981 &$      8.09 $&$      2.97 $&1.12&$  1 $\\
 2457186.79015 &$      0.48 $&$      2.84 $&13.94&$  1 $\\
 2457186.82363 &$      6.96 $&$      3.18 $&0.83&$  1 $\\
 2457187.67459 &$      5.04 $&$      1.66 $&12.95&$  1 $\\
 2457187.74382 &$      2.94 $&$      2.41 $&19.73&$  1 $\\
 2457187.78138 &$      0.31 $&$      2.22 $&11.51&$  1 $\\
 2457187.81506 &$      8.21 $&$      3.13 $&9.47&$  1 $\\
 2457188.63849 &$      6.46 $&$      2.19 $&16.38&$  1 $\\
 2457188.69403 &$      5.69 $&$      1.73 $&5.09&$  1 $\\
 2457188.75958 &$     -8.45 $&$      3.83 $&17.25&$  1 $\\
 2457188.79906 &$     -1.28 $&$      2.62 $&10.10&$  1 $\\
 2457188.83141 &$     10.79 $&$      3.66 $&-8.32&$  1 $\\
 2457189.64108 &$     -1.58 $&$      2.19 $&-9.69&$  1 $\\
 2457189.69856 &$     -1.46 $&$      1.71 $&22.02&$  1 $\\
 2457189.75643 &$     -3.27 $&$      1.53 $&9.36&$  1 $\\
 2457189.78890 &$     -8.02 $&$      1.53 $&32.01&$  1 $\\
 2457189.82294 &$     -8.48 $&$      1.78 $&7.80&$  1 $\\
 2457190.64915 &$     -6.05 $&$      1.67 $&4.34&$  1 $\\
 2457190.68400 &$     -2.64 $&$      1.73 $&19.98&$  1 $\\
 2457190.71716 &$     -7.75 $&$      2.04 $&21.29&$  1 $\\
 2457190.74988 &$     -3.93 $&$      2.18 $&6.21&$  1 $\\
 2457190.78420 &$     -8.97 $&$      2.22 $&4.07&$  1 $\\
 2457191.66878 &$    -10.48 $&$      3.27 $&0.45&$  1 $\\
 2457191.70642 &$    -13.28 $&$      2.14 $&24.39&$  1 $\\
 2457191.76095 &$     -4.02 $&$      2.42 $&-2.81&$  1 $\\
 2457191.79493 &$     -4.92 $&$      3.20 $&22.16&$  1 $\\
 2457191.82491 &$     -0.74 $&$      3.98 $&14.21&$  1 $\\
 2457192.65964 &$     -2.13 $&$      3.69 $&13.14&$  1 $\\
 2457198.72143 &$     -2.16 $&$      4.51 $&8.10&$  1 $\\
 2457198.80982 &$    -11.76 $&$      2.74 $&11.57&$  1 $\\
 2457199.60136 &$     -9.32 $&$      1.89 $&15.52&$  1 $\\
 2457199.68346 &$    -13.58 $&$      1.59 $&4.78&$  1 $\\
 2457211.70008 &$     11.06 $&$      2.31 $&19.41&$  1 $\\
 2457253.48543 &$     -0.60 $&$      2.50 $&8.17&$  1 $\\
 2457264.51816 &$      6.03 $&$      2.99 $&-6.49&$  1 $\\
 2457267.56388 &$     15.81 $&$      2.15 $&15.49&$  1 $\\
 2457270.50147 &$     -0.73 $&$      1.92 $&7.98&$  1 $\\
 2457278.56859 &$     12.00 $&$      3.48 $&-4.74&$  1 $\\
\hline
\end{tabular}
\tablecomments{0: PFS; 1: HARPS}
\label{rv_table}
\end{table}

\end{document}